\documentstyle[psfig]{mn}
\newcommand{\bib}{\bibitem[\protect\citename}
\newcommand{\egrs}{extragalactic radio source }
\newcommand{\lang}{Laing-Garrington }
\newcommand{\lp}{Liu-Pooley }
\newcommand{\uv}{{\sl uv}}

\newcommand{\hsb}{high surface brightness }
\newcommand{\etal}{{et al.}}

\begin{document}
\title[Jets, lobe length \& spectral index]{Asymmetry of jets, lobe length
and spectral index in quasars}
\author[J. Dennett-Thorpe et al.]
   {J. Dennett-Thorpe,$^1$\thanks{jdt@sintra.cc.fc.ul.pt}
     A.~H. Bridle,$^2$ P.~A.~G. Scheuer,$^1$ \cr
    R.~A. Laing $^3$ and J.~P. Leahy $^4$ \\
     $^1$ Mullard Radio Astronomy Observatory, 
     Cavendish Laboratory, Madingley Road, Cambridge CB3 0HE \\ 
     $^2$ National Radio Astronomy Observatory, 
     520 Edgemont Road, Charlottesville VA 22903-2475, USA  \\
     $^3$ Royal Greenwich Observatory, Madingley Road, 
     Cambridge CB3 0EZ  \\
     $^4$ Nuffield Radio Astronomy Observatories, Jodrell Bank, Macclesfield, 
     Cheshire SK11 9DL }

\date{Received }
\maketitle

\begin{abstract}
The less depolarized lobe of a radio source is generally the lobe
containing the jet (Laing-Garrington correlation) but the less
depolarized lobe is also generally that with the flatter radio
spectrum (Liu-Pooley correlation). Both effects are strong; taken
together they would imply a correlation between jet side and lobe
spectral index, i.e. between an orientation-dependent feature and one
which is intrinsic. We test this prediction using detailed spectral
imaging of a sample of quasars with well-defined jets and investigate
whether the result can be reconciled with the standard interpretation
of one-sided jets in terms of relativistic aberration. Our central
finding is that the spectrum of high surface brightness regions is
indeed flatter on the jet side, but that the spectrum of low surface
brightness regions is flatter on the side with the longer lobe. We
discuss possible causes for these correlations and favour
explanations in terms of relativistic bulk motion in the high surface
brightness regions and differential synchrotron ageing in the extended
lobe material.
\end{abstract}


\section{Introduction}
\subsection{The problem}
Laing, Garrington and colleagues \cite {gnat,ral} showed that the lobe
of a powerful radio source containing the jet is almost invariably
less depolarized than its counterpart on the opposite side of the
nucleus; 39 out of the 47 sources in Tables~3 to 6 of Garrington,
Conway \& Leahy \shortcite{gcl} obey the Laing--Garrington rule
(only 4 sources definitely show the reverse effect).
These 47 sources all have FRII morphologies \cite{FR} and most are
quasars, although a few radio galaxies are included.  Very few FRII
sources have detectable jets on both sides of the nucleus, and we will
use the term ``the jet'' to refer to the brighter (in almost all cases
the only) such feature.  The most widely accepted explanation of the
Laing--Garrington effect is that jets are intrinsically two-sided and
relativistic: the nearer one appears brighter as a result of Doppler
beaming and radiation from its lobe passes through less of the
depolarizing medium around the source \cite{ral,gar91}. Thus the
Laing--Garrington effect is explained as a consequence of orientation.

Liu \& Pooley \shortcite{lpa} found that there is also a strong
correlation between depolarization and lobe spectral index: the lobe
with the flatter spectrum is the less depolarized. Their original
sample contained a majority of sources without detectable jets, but 33
of the 47 sources with strong jets in Tables~3 to 6 of Garrington {et
al.} also obey the Liu--Pooley rule.

Taken together these two strong correlations imply that the lobe spectrum is 
flatter on the jet side of a source. According to the standard model of an 
\egrs  the lobe material is almost static relative to the host galaxy and 
therefore any motion of the lobe material is inadequate to account for 
significant differences between the lobe spectra as an orientation-related 
effect. Thus the two correlations together constitute a {\it prima facie} case 
against the standard model of `Doppler-boosted' relativistic jets. 

The most plausible defence is that intrinsic and orientation effects both
operate:
\begin{enumerate}
\item The \lang sample includes only sources with detected (and, in most
cases, prominent) jets.  The majority are identified with quasars, which
tend to have much brighter jets than radio galaxies of the same 
total power.   By contrast, the \lp sample was not selected on 
jet emission, and consists chiefly of powerful radio galaxies. 
\mbox{According} to `unified theories' of radio galaxies and quasars 
\cite{Sch87,Bar89}, all of the sources form part of the same population, being
classified as quasars if their jet axes are closer than $\approx$50$^\circ$
to the line of sight, otherwise as radio galaxies.  Faraday rotation and
depolarization increase with the amount of ionized gas and magnetic field
along the line of sight to the emitting region, and differences between the two
lobes will result either from orientation or from intrinsic asymmetries in
the surrounding material. Differences between the path lengths to the two
lobes will be much larger for the quasars, and we might expect orientation
effects to dominate.  For the radio galaxies, on the other hand,
path-length differences should be small, and intrinsic effects will be
relatively more important.
\item For radio galaxies without strong jets, the shorter lobe is also the
more depolarized, but this effect is barely significant in quasars
\cite{L93}, despite the fact that the nearer lobe should appear longer as
a result of differential light-travel effects \cite{ryl65}.
The shorter lobe in radio galaxies is also associated
with brighter line-emitting gas \cite{MvBK91}, and this cannot be an
orientation effect.  We therefore require an intrinsic mechanism which
relates lobe length, spectral index and depolarization, and which dominates
in radio galaxies, together with an orientation-dependent mechanism which
relates jet sidedness, spectral index and depolarization for quasars.
\item The most plausible explanation for a correlation between spectral
index and jet sidedness is that the emission from a lobe might include
significant contributions from the jet or an associated hotspot, both 
with flatter spectra than the surrounding material \cite{gcl,tribble}.
It is widely believed that the flow velocity of hotspot material is a
substantial (though ill-defined) fraction of the speed of light
\cite{bbr,bp84} and therefore that significant Doppler beaming should
occur in hotspots.
\end{enumerate}

If unified models are correct, then the intrinsic mechanism postulated for
radio galaxies must also operate in quasars, in competition with orientation
effects.  Evidently what is needed is the direct observation of the
spectral indices of the lobes of radio sources with jets, with enough
resolution to discriminate clearly between lobe, jet and hotspot. 
In this paper we report detailed comparisons of the spectral index
distributions in a small sample of quasars; a preliminary account based on the
sources analysed at that time was given at the Mt. Stromlo symposium of
1993 \cite{blst}. (Note, however, that Fig.~1 of that paper is
incorrectly drawn: some sources have jet and counter-jet side labels reversed.)

\subsection{Previous work}
Garrington {et al.} (1991) mapped 47 quasars with jets, at 1.4 and
5~GHz, and found that in 37 out of 47 the side of the source with the
jet had the flatter spectrum (see table~3 of that paper). These
observations, then, show the direct correlation which challenges the
standard model of `Doppler-boosted' jets. However, as the authors
themselves state, the images do not have enough angular resolution to
permit further investigation of the causes of the correlation; in
particular, in most cases they do not adequately resolve the hotspots
from the relatively low-brightness lobes to allow the spectral index
of the lobes to be measured reliably.

Barthel {et al.} \shortcite{pb88} and Lonsdale \etal\ 
\shortcite{pb93} made images of over 100 quasars, many of which were
observed at both 5 and 15~GHz and also have clearly detected jets. So
far as we know, these have not been investigated from the point of
view of the present paper, and it is not clear that these data will
lend themselves to measuring the distribution of spectral index in
low-brightness regions.

Lonsdale and Morison \shortcite{lon83} find spectral asymmetries in
the hotspots of four powerful radio sources. In two of these sources
jets have now been detected (3C268.4 and 3C249.1; the latter is also
in our sample); in these sources the hotspot spectrum is flatter on
the jet side. In the other two sources the flatter spectrum is found
in the more compact component, which is preferentially found on the
jet side \cite {lai89,bhlbl}.

Throughout this paper, we define the spectral index $\alpha$ in the sense
flux density $\propto$ frequency$^{-\alpha}$.
 
\section{Observations}
\subsection{The sample}
Because we wish to explore spectral index distributions, we are
restricted to a small sample of sources which are bright enough to be
mapped in detail. We therefore started with the 12 quasars of which
Bridle {et al.} (1994a, hereafter BHLBL) had already made detailed
5~GHz images, plus 3C47 \cite{fernini} which satisfies the same
selection criteria and for which 5~GHz data of similar quality were
available. The BHLBL sample was a subset of the 19 brightest quasars
with angular size greater than 10 arcsec in the 3CR catalogue, the
only further selection being for reasons of scheduling.  Of these 13
quasars, we selected those with prominent one-sided jets and fairly
standard appearance; thus we excluded 3C68.1 (ambiguous jet
sidedness), 3C215 (jets on both sides of the core, and $90^\circ$
distortion between smallest and largest scales) and 3C9 (because there
is little of the source that can be described unequivocally as `lobe'
rather than `jet').

\subsection{Observing programme}
In order to derive spectral-index maps, we made new observations at
1.4 and 1.7~GHz with {\small MERLIN} and the National Radio Astronomy
Observatory (NRAO) {\small VLA}, and extracted
additional data from the {\small VLA} archive. The 1.4--1.7~GHz
observations were designed to provide as much overlap as possible in
\uv~coverage with the 5~GHz observations of BHLBL.  Most sources were
just observed with the {\small VLA} in the {\small A} configuration,
but 3C208 and 3C432 are only 14.6 and 14.8 ~arcsec in extent,
respectively, so that even the {\small VLA~A~}array does not provide
enough angular resolution; for these, {\small MERLIN} observations
were obtained and combined with the {\small VLA} data. Observations of
3C351 at 1417.5~MHz in {\small A} array poorly covered the \uv~plane,
and  undersampled the large scale structure. For this reason the data
were combined with {\small B} array observations.  Table
\ref{tab:obs} shows a summary of the lower-frequency observations.

\begin{table}
\vspace{5mm}
\caption{Observing schedule}
 \begin{tabular}{p{0.75cm}lllc}\hline
 Source & Observing & Duration & Array & Frequency 
  \\
  & date & & & (MHz)\\ \hline
 3C47  & 1992 Dec 7  & 399m& VLA-A & 1664.9, 1635.1 \\
       & 1986 Jul 12 & 36m & VLA-B & 1652.4\\
 3C175 & 1992 Oct 31& 89m & VLA-A & 1417.5, 1467.5 \\
 3C204 & 1982 Mar 1 & 18m & VLA-A & 1417.5\\
 3C208 & 1981 Feb 24& 19m & VLA-A & 1464.9 \\
       & 1994 Aug 14& 14h & MERLIN & 1420.0, 1658.8 \\
 3C263 & 1992 Oct 31& 89m & VLA-A & 1417.5, 1467.5 \\
 3C249.1 & 1982 Mar 1& 38m & VLA-A & 1417.5\\
 3C334 & 1992 Oct 31 & 138m& VLA-A & 1417.5, 1467.5 \\
 3C336 & 1992 Oct 31 & 47m & VLA-A & 1417.5, 1467.5 \\
 3C351 & 1982 Mar 11 & 12m & VLA-A & 1417.5\\
       & 1982 May 30 & 26m & VLA-A & 1417.5\\
       & 1987 Nov 25 & 23m & VLA-B & 1452.4,1502.4\\
       & 1987 Dec  2 & 6m  & VLA-B & 1452.4,1502.4\\
 3C432 & 1992 Nov 1& 90m & VLA-A & 1464.9, 1514.9 \\
  & 1994 Aug 1& 14h & MERLIN & 1420.0, 1658.8\\
\hline
\end{tabular}
\label{tab:obs}
\end{table}

The {\small VLA} data were reduced and calibrated using the NRAO
{\small AIPS} package. Initial editing and calibration of {\small MERLIN} data
was performed using the {\small OLAF} package at Jodrell Bank,
before being transferred to {\small AIPS} for self-calibration and
imaging. Where possible only phase self-calibration was used.  In a few cases with significant amplitude errors (for
example when combining VLA and MERLIN data; see below) amplitude
self-calibration was used with a long integration interval.

In order to maximise the \uv~coverage of the sparse {\small MERLIN}
array, multi-frequency synthesis imaging was used \cite{Conway90}. The
data from the two frequencies were first corrected in total flux for
the spectral index of the source and then combined to form a single data
set. The spectral index correction was calculated using the compact
components from the 5~GHz and one of the 1.4 or 1.7~GHz images. The
dominant spectral errors will come from bright compact components
(hotspots, core), yielding an error of $\approx I \alpha^\prime / 200$
near the hotspots, where $I$ is the peak surface brightness and
$\alpha^\prime$ is the residual spectral
index, i.e. the difference between the spectral index assumed for the
purposes of correction and the true spectral index of the compact
feature \cite{Conway90}.  The spectral errors will be below the level
due to thermal noise and dynamic range limitations if the residual
spectral index of the compact component is $\la$~0.15.  The residual
spectral index is less than 0.15 near the hotspots; and in both sources
imaged by this method, there is no lobe
emission near the core, where larger errors might occur. Our images
are therefore limited by `reconstruction errors', not spectral errors and
thus we are confident in using these images for the purposes of
spectral index analysis.  The {\small MERLIN} data were combined with
the {\small VLA} data with weights proportional to the inverse
expected rms noises on a single integration point. The combined data
were then further calibrated, including a few passes of amplitude
self-calibration.

Images at all frequencies were made using the {\small AIPS CLEAN}
algorithms {\small APCLN} and {\small MX}. All images were {\small
CLEAN}ed to the noise level to ensure that all significant emission,
especially in the low brightness regions, was restored with the same
effective resolution by {\small CLEAN}.

\subsection{The 1.4~GHz images}

\begin{figure*}
%%\vspace{25cm}
\centerline{
\psfig{figure=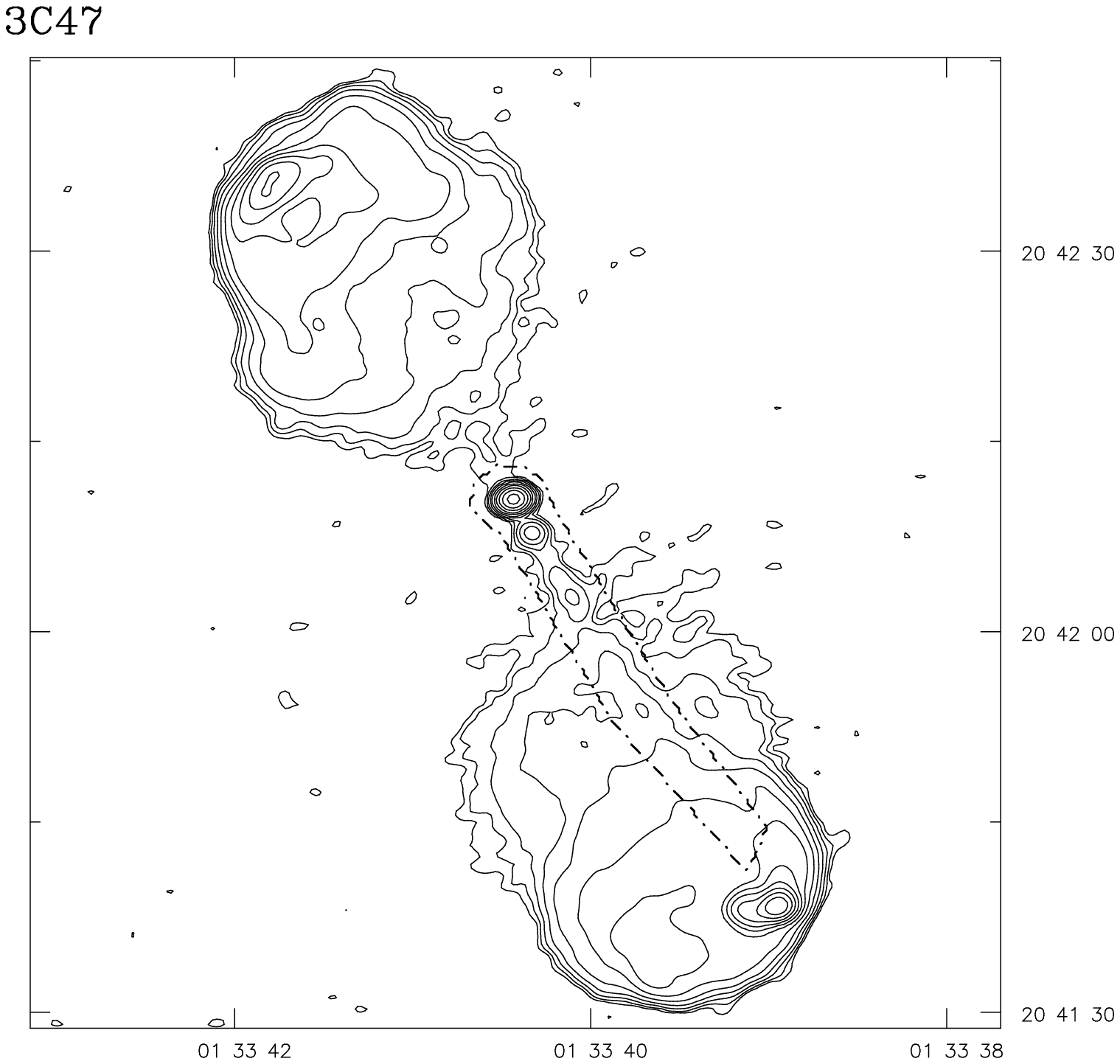,angle=0,height=5cm,clip=}
\psfig{figure=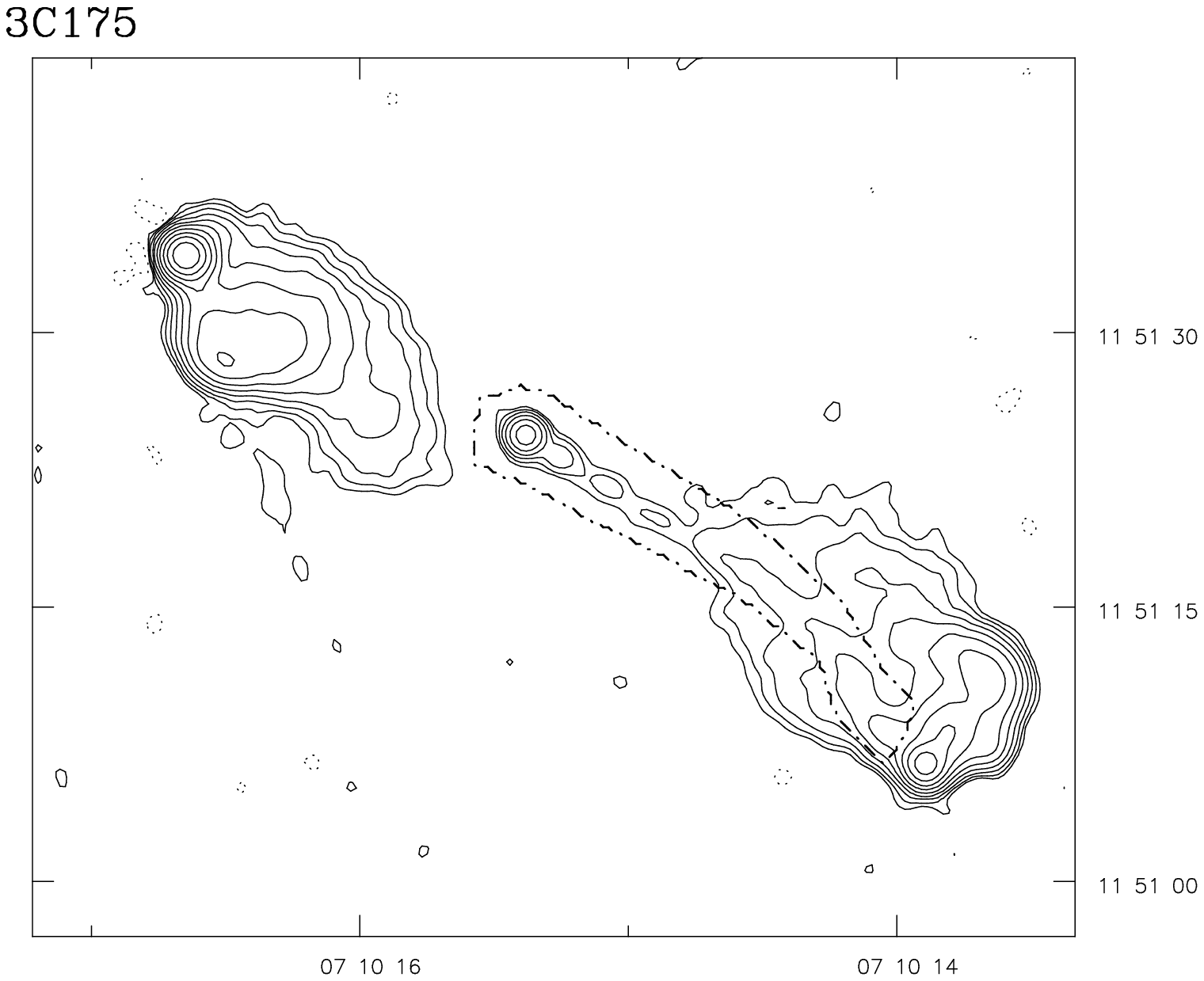,angle=0,height=5cm,clip=}
}

\centerline{
\psfig{figure=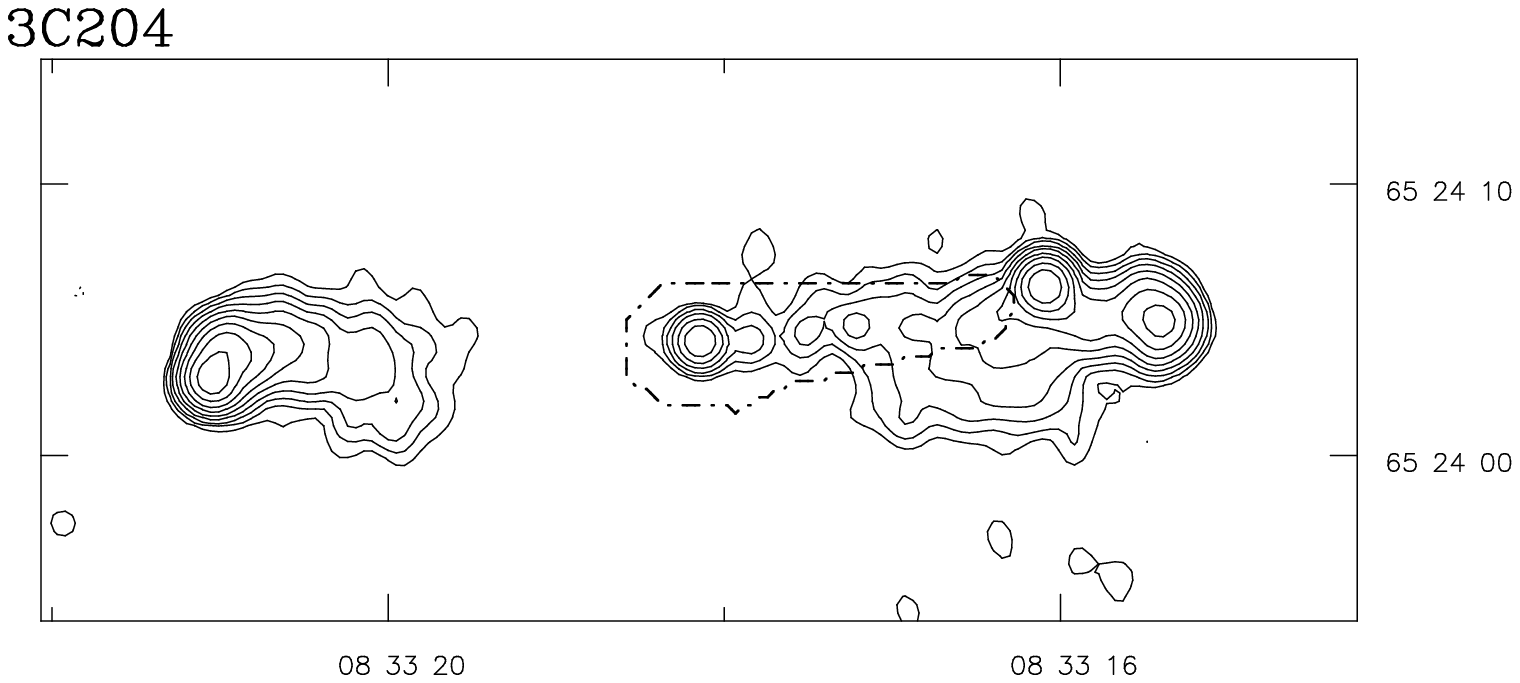,angle=0,height=4cm,clip=}
\psfig{figure=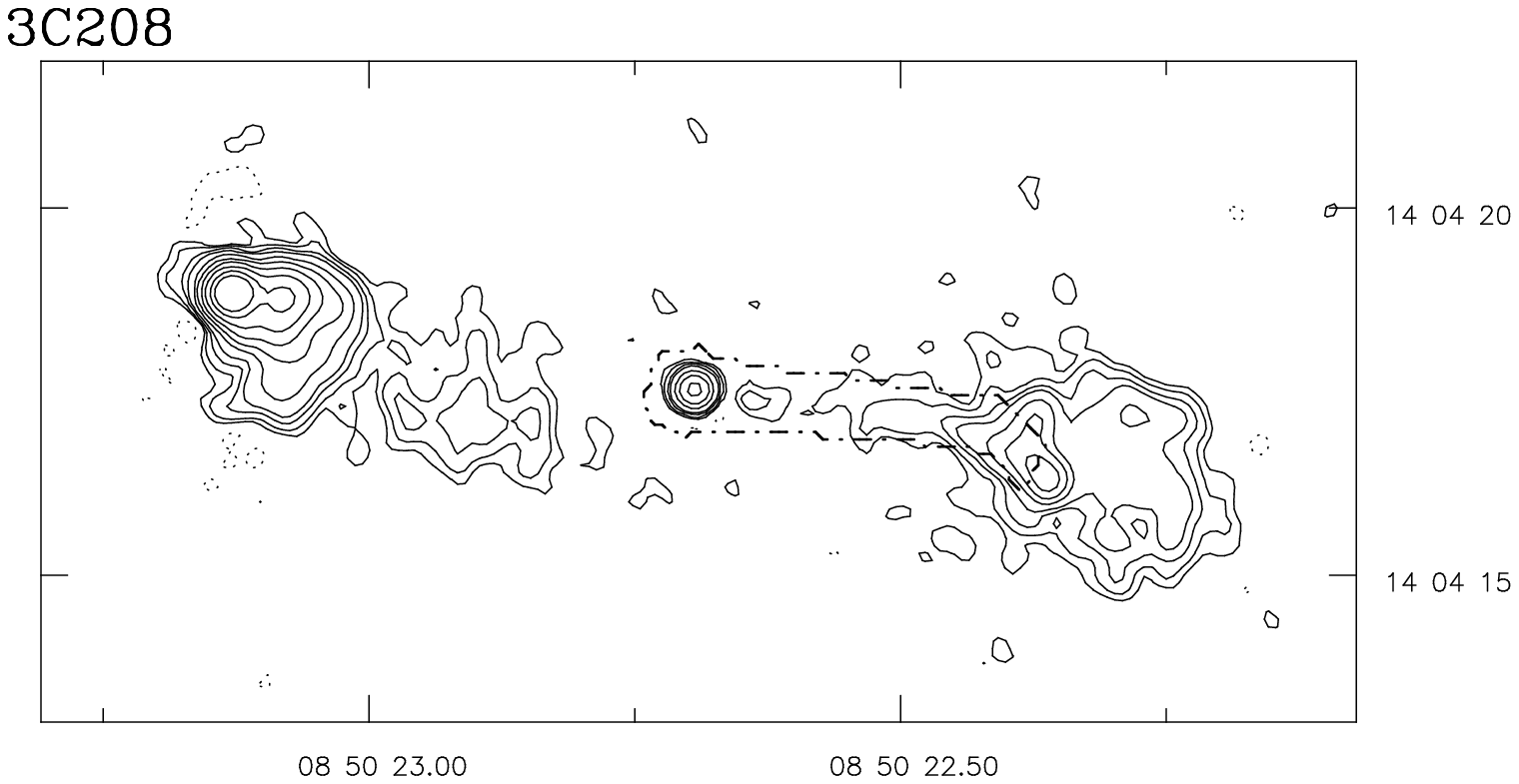,angle=0,height=4cm,clip=}
}

\centerline{
\psfig{figure=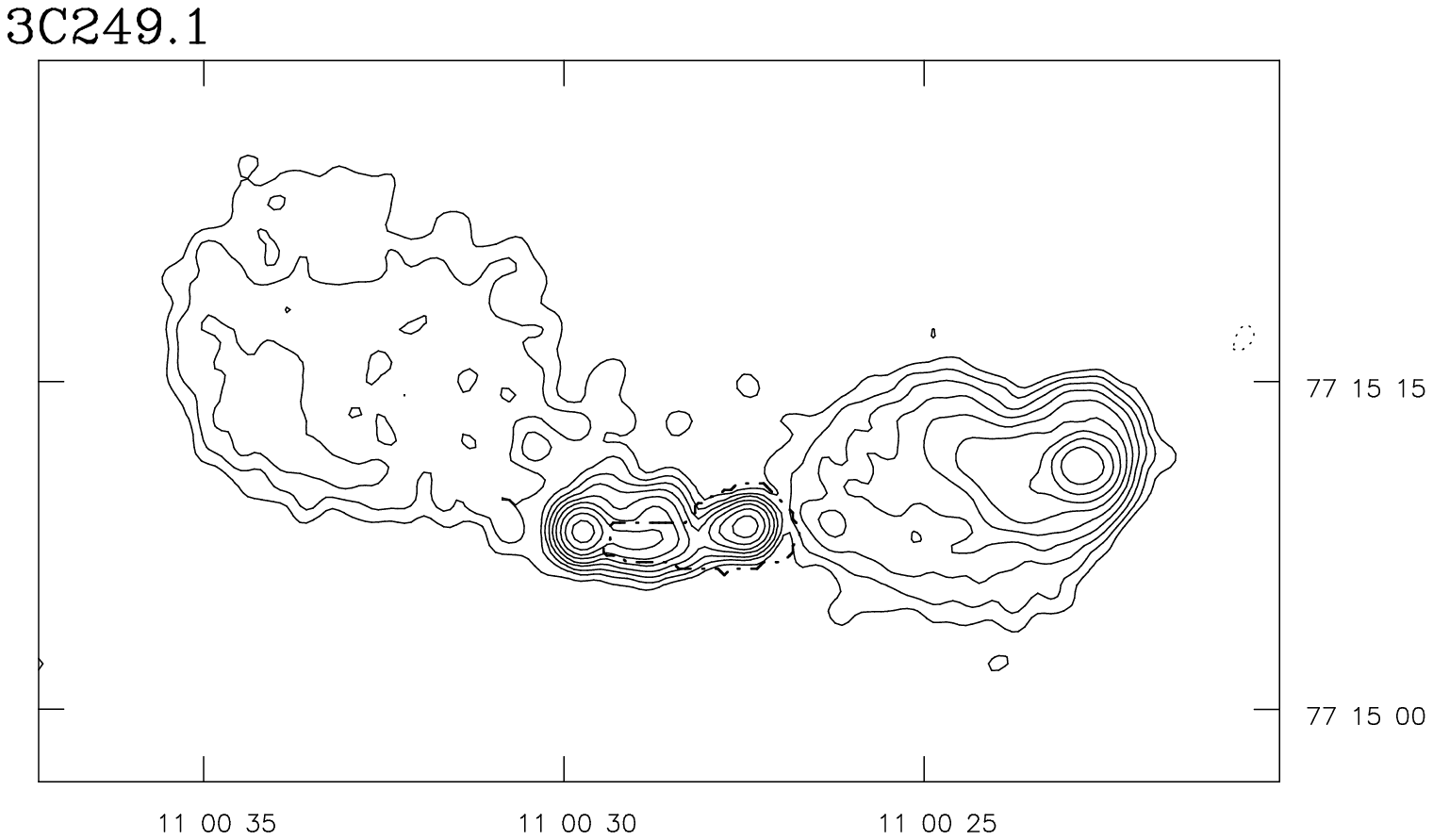,angle=0,height=4.6cm,clip=}
\psfig{figure=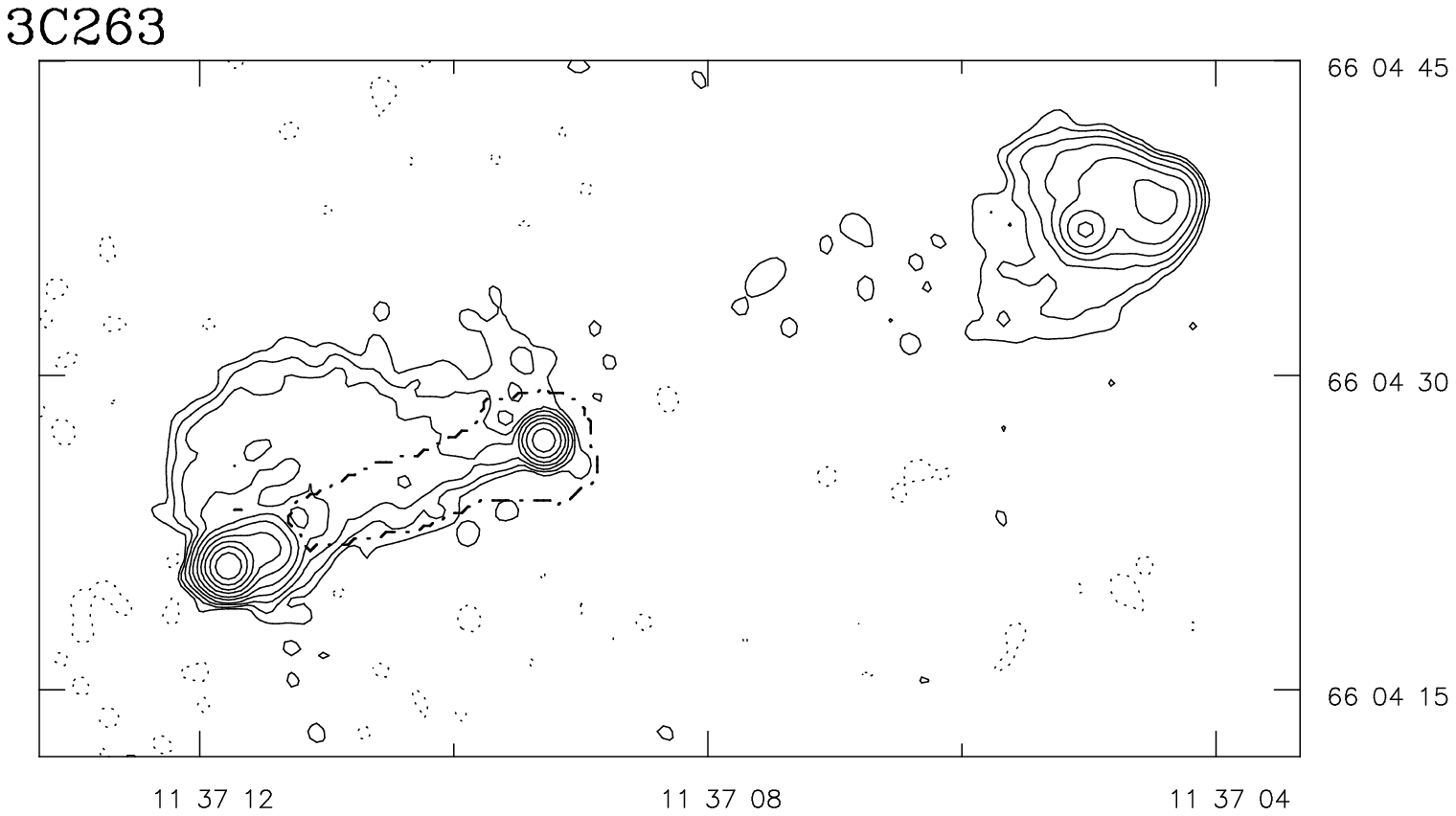,angle=0,height=4.6cm,clip=}
}

\centerline{
\psfig{figure=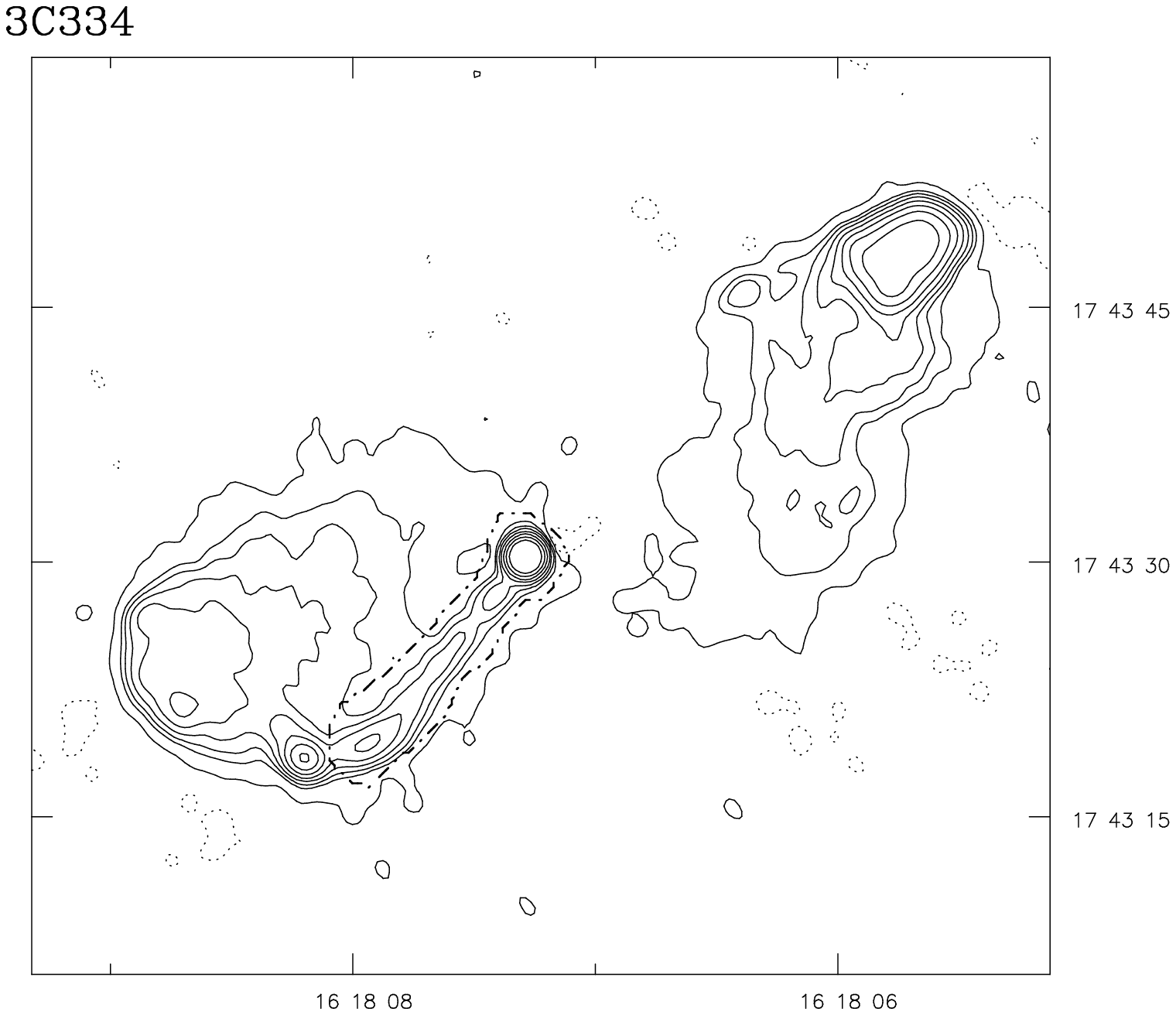,angle=0,height=5cm,clip=}
\psfig{figure=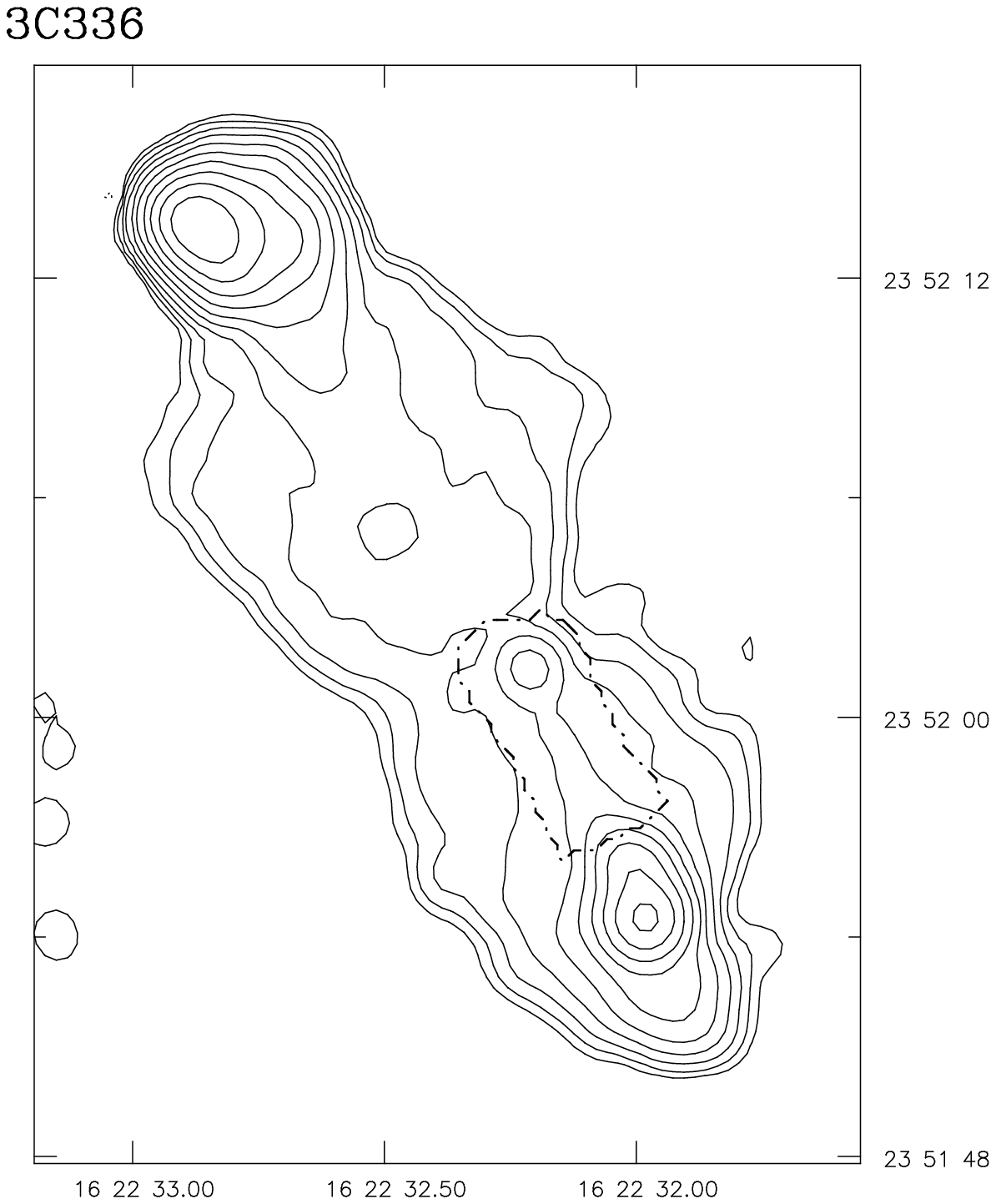,angle=0,height=5cm,clip=}}

\centerline{
\psfig{figure=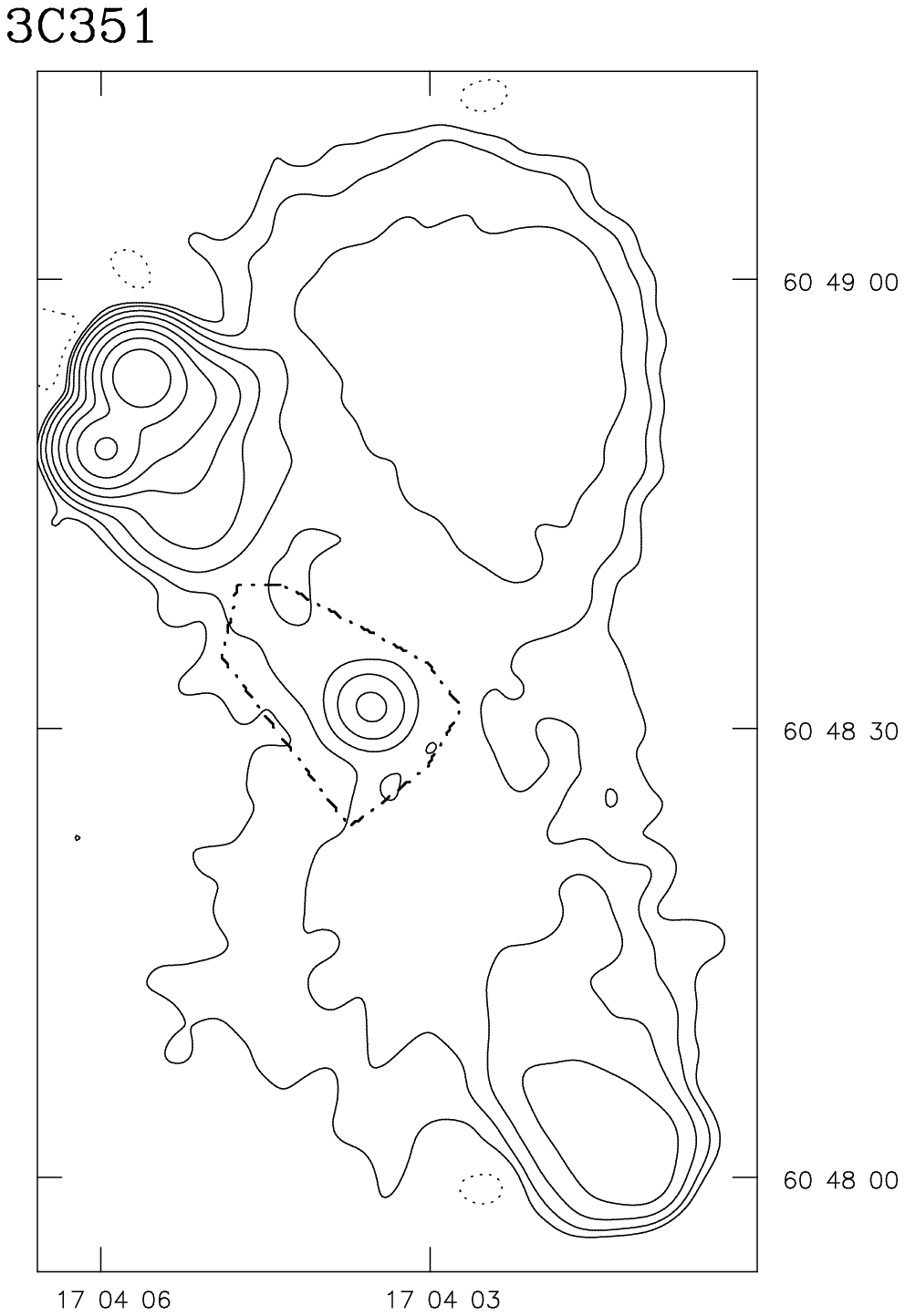,angle=0,height=4.6cm,clip=}
\psfig{figure=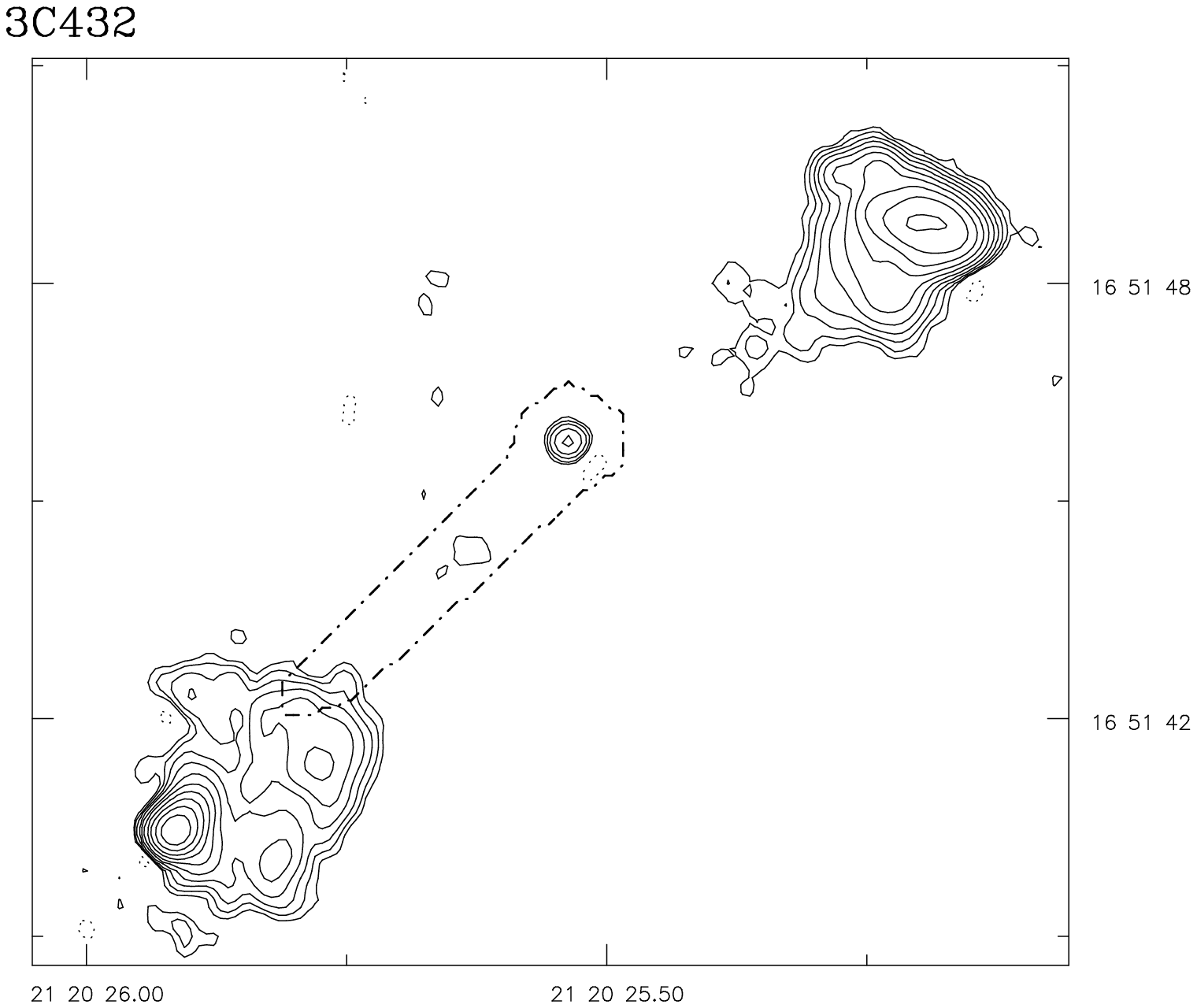,angle=0,height=4.6cm,clip=}
}
\caption{The 1.4/1.7~GHz images of the sample. The contours are those used
to divide the sources into surface brightness regions and the dotted lines
enclose the core and jet regions excluded from the spectral
index calculations}
\label{maps}
\end{figure*}

The images are shown in Fig.~\ref{maps}. The contours are
those used to divide the sources up into zones for the spectral index
comparisons (see below): the lowest solid contour is at 3$\sigma$ and
the others are evenly spaced in $\log$(surface brightness).

Table~\ref{params} shows the imaging parameters used and the total
flux density in the images to the 3$\sigma$ contour on the 1.4/1.7~GHz
image. The single dish total flux densities used were the 1400, 2695 and
5000~MHz flux densities of Laing \& Peacock \shortcite{lai80},
interpolated to the observing frequency. 
Our L-band image of 3C334 apparently contains only 87 per cent of the single dish 
flux density, but the latter contains a contribution from another source 
of 114mJy 4.4~arcmin from 3C334 itself (as noted by BHLBL). Precisely how much that 
source contributed depends on details of how the single dish measurements 
were analyzed to allow for finite size, but it seems likely that most of the 
flux of the confusing source was included. Our image then contains 92
per cent of the 
corrected single-dish flux density. It is seen from Table 2 that
we have a good representation of the total source flux density in all
our other images. 3C351 has been imaged at a lower resolution than the other
sources in order to show the extended region of low surface brightness
preceding the N hotspots as this region was poorly represented at
higher resolution.

\begin{table*}
\vspace{5mm}
\caption{Image parameters}
\begin{tabular}{llrrrrrrr} \hline\\
& & \multicolumn{4}{c}{1.4/1.7~GHz} & \multicolumn{3}{c}{4.86~GHz} \\
       & FWHM           & noise & contour& total flux & \% single
       dish& noise     &total flux & \% single dish\\
       &                & ($\mu$Jy/bm)& factor& density (Jy)     &
       flux density&\footnotesize{($\mu$Jy/bm)}&  density (Jy)  & flux density\\
\hline
3C47   &1.45 \sf{x} 1.13& 48    & 1.96 &3.36    &93  & 35& 1.22 & 111\\  
3C175  &1.30            & 151   & 2.01 &2.53    &105 & 38& 0.67 & 100\\ 
3C204  &1.15            & 199   & 1.91 &1.19    &89  & 61& 0.32 & 93\\  
3C208  &0.35            & 250   & 1.92 &2.16    &104 & 25& 0.56 & 102\\ 
3C249.1&1.20            & 242   & 1.87 &2.33    &104 & 32& 0.81 & 101\\ 
3C263  &1.10            & 197   & 2.20 &3.15    &103 & 57& 1.11 & 105\\ 
3C334  &1.30            & 73    & 1.80 &1.83    &87  & 36& 0.63 & 106\\ 
3C336  &1.25            & 187   & 2.00 &2.63    &99  & 38& 0.82 & 115\\ 
3C351  &3.00            & 222   & 3.49 &3.13    &90  & 105& 1.18 & 96\\
3C432  &0.37            & 234   & 1.76 &1.46    &104 & 19& 0.37 & 114\\
\hline
\end{tabular}
\label{params}
\end{table*}

\subsection{The comparison between 1.4~GHz and 5~GHz images}
As stated in the Introduction, we wish to compare the spectra of the
lobes on the jet and counterjet sides, taking care to distinguish
between lobes, hotspots and jets. Nevertheless we begin by noting
that, when we compare the entire contents of the lobes, as in
Garrington et al.(1991), the correlation found by these authors also
appears in our data. Table \ref{tab:total-spix} shows that in 7 cases
out of 10 the jet side has the flatter spectrum, in two there is no
significant difference and in one the jet side has the steeper
spectrum. Thus the spectra of our sample are not atypical.

As soon as we try to compare the spectra of lobes (excluding jets and 
hotspots) a fundamental complication becomes obvious: there is no such thing 
as `the spectral index of the lobe'.
The spectrum steepens progressively (though usually not very
regularly) from the neighbourhood of the hotspot towards the middle of
the source (a trend generally attributed to synchrotron losses). Some
scheme must be invented for comparing like with like on the two
sides. One might consider comparing spectral indices at the same
fractional distance from the core to the end of the lobe (or to the
hotspot); that is not particularly satisfactory in practice as the
lobes are often morphologically very different, they may have
protrusions and the hotspots are often recessed from the end of the
lobe.  The scheme we have adopted is to compare regions with the same
surface brightness.  That brings certain 
complications with it (see e.g. Section 4.2) but other schemes that we 
considered were much less satisfactory, and the adopted scheme at least has 
the merit of leading to an interesting result (Section 3). 
The analysis therefore proceeds as follows:

\begin{table}
\begin{center}
\caption{Total lobe spectral indices, determined from the ratio of
flux densities inside the 3$\sigma$ contour on the 1.4/1.7~GHz image and the
identical region in the 5~GHz image. Cores have been excluded.
}
\begin{tabular}{lrr}
\hline
Source & $\alpha^{js}$ & $\alpha^{cjs}$ \\
\hline
3C47   & 0.85 & 0.94\\
3C175  & 1.07 & 1.15\\
3C204  & 1.14 & 1.19\\
3C208  & 1.14 & 1.25\\
3C249.1& 0.86 & 0.96\\
3C263  & 0.96 & 0.88\\
3C334  & 0.99 & 1.00\\
3C336  & 0.96 & 0.98\\ 
3C351  & 0.78 & 0.88\\
3C432  & 1.05 & 1.28\\
\hline
\multicolumn{3}{c}{(js) refers to jet side;} \\
\multicolumn{3}{c}{(cjs) to counter-jet side.}\\
\end{tabular}
\label{tab:total-spix}
\end{center}
\end{table}

\begin{figure}
%%\vspace{7cm}
\centerline{
\psfig{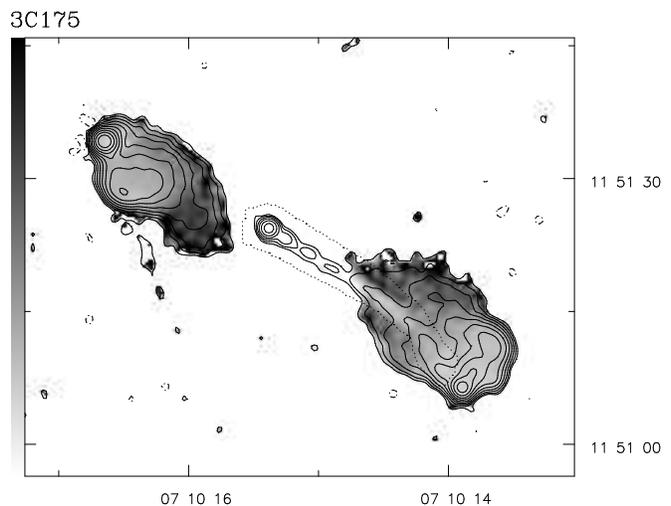}
}
\caption{Subdivision of 3C175 into surface brightness zones. The
contours are spaced as for the spectral index calculations on the
1.4~GHz image and the grey scale represents the spectral index between
1.4 and 5.0~GHz. The dotted line again shows the excluded jet and core
region.  Contours : -.465, .465, .931, 1.865, 7.479, 14.96, 30.00,
60.08, 120.3, 241.0 mJy/beam. Grey scale: 0.6 (white) -- 2.5 (black) }
\label{zones}
\end{figure}

\begin{enumerate}
\item Make 5~GHz and 1.4/1.7~GHz images at the same angular resolution, using 
as near as possible the same \uv~coverage; the authors of BHLBL kindly 
made the calibrated 5~GHz visibilities available. Correct the images for any
zero-level offsets using estimates of the mean off-source level.
\item Cut out the region of the jet and the core; the excluded regions are 
indicated in Fig.~\ref{maps}. 
\item Subdivide the resulting images into regions within fairly narrow
ranges of surface brightness at 1.4 or 1.7~GHz. The ranges of surface
brightness were chosen as follows: the lowest contour was set at
three times the rms noise level on the 1.4 or 1.7~GHz image, and the
image was divided into logarithmically equally spaced contours. Each
source was divided into about 10 zones; the precise number depends on
the dynamic range and was chosen so that each zone contains a
sufficiently large number of beams to ensure that the spectral error
introduced by the thermal noise is small, and so that the zone is
generally wider than a beam. The contours shown in
Fig.~\ref{maps} are the contours dividing the zones. Fig.~\ref{zones}
shows the same contours for 3C175, but with a grey scale of spectral
index superposed. This demonstrates that surface brightness and
spectral index are roughly correlated.
\item Compute the spectral index for each surface brightness range on
each side of the source from the ratio of the total flux densities in
that surface brightness range at 5~GHz and 1.4/1.7~GHz.  
Every effort was made to remove any small zero-level offsets from the
images but note that the {\em differences} between the spectra on the two
sides of a source, at the same surface brightness, are in any case
insensitive to such errors, and are unaffected by inaccuracies in the
flux scale.
\end{enumerate}

\section{Results}

\begin{figure*}
%%\vspace{25cm}

\centerline{
\psfig{figure=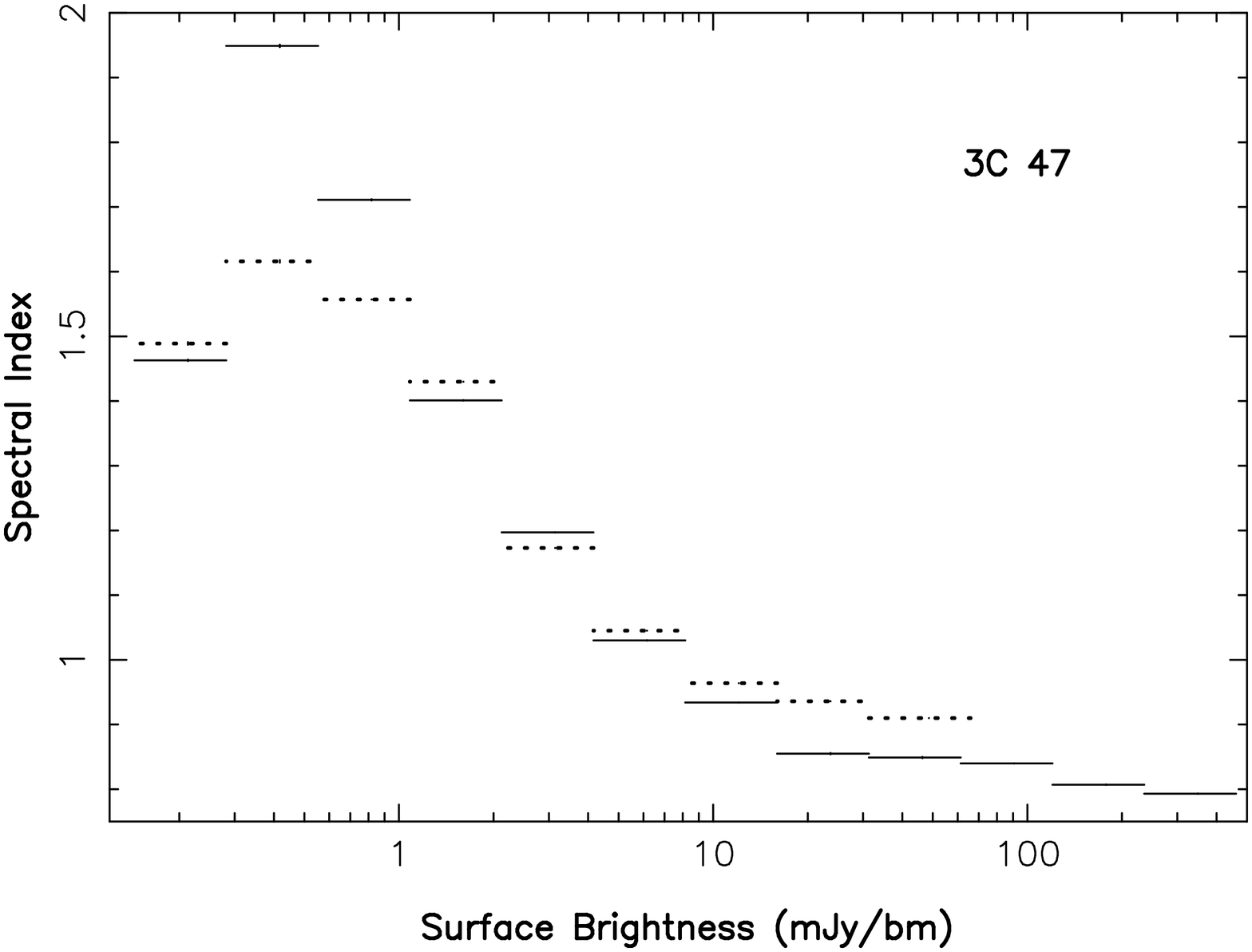,angle=-90,height=4.6cm,clip=}
\psfig{figure=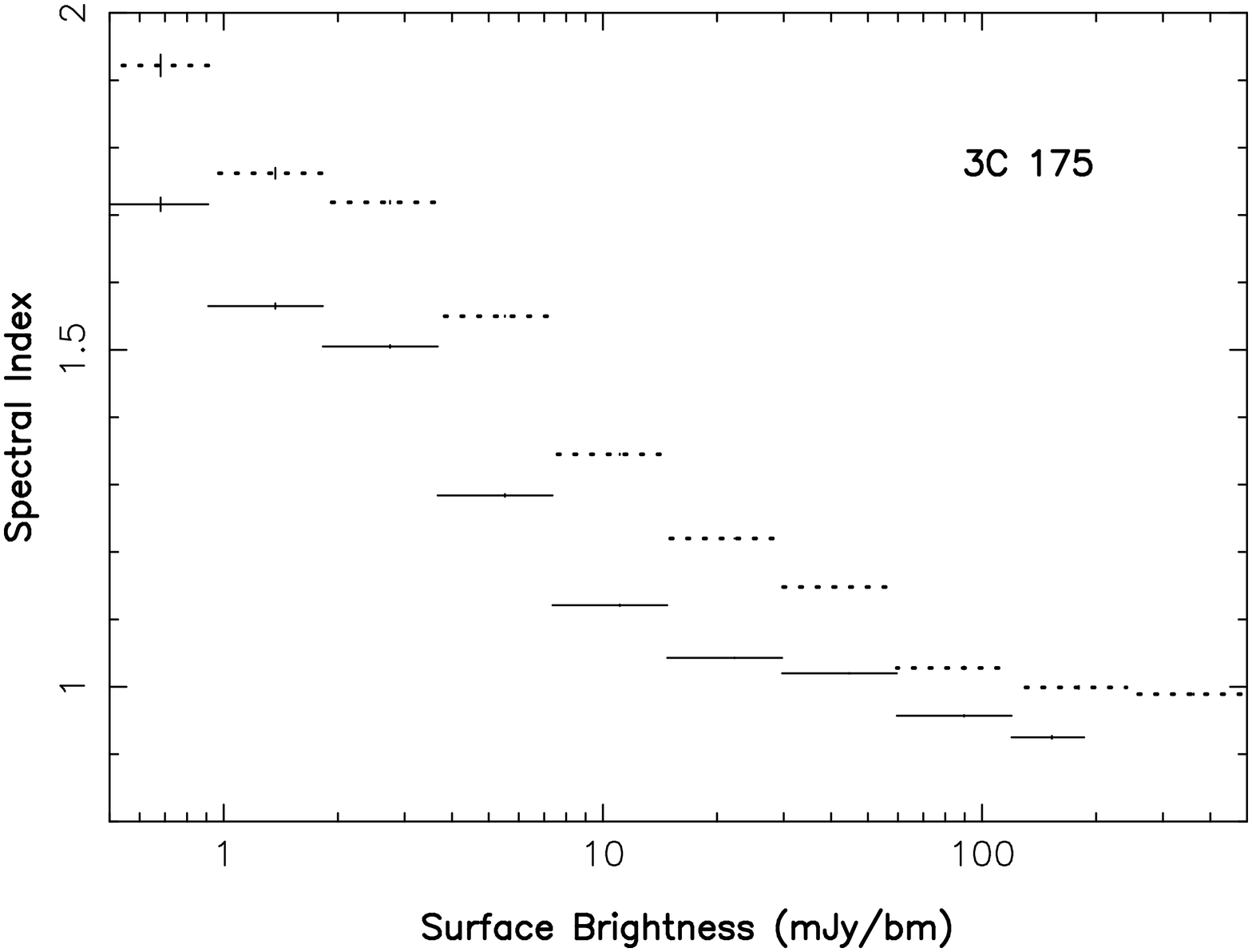,angle=-90,height=4.6cm,clip=}
}
\centerline{
\psfig{figure=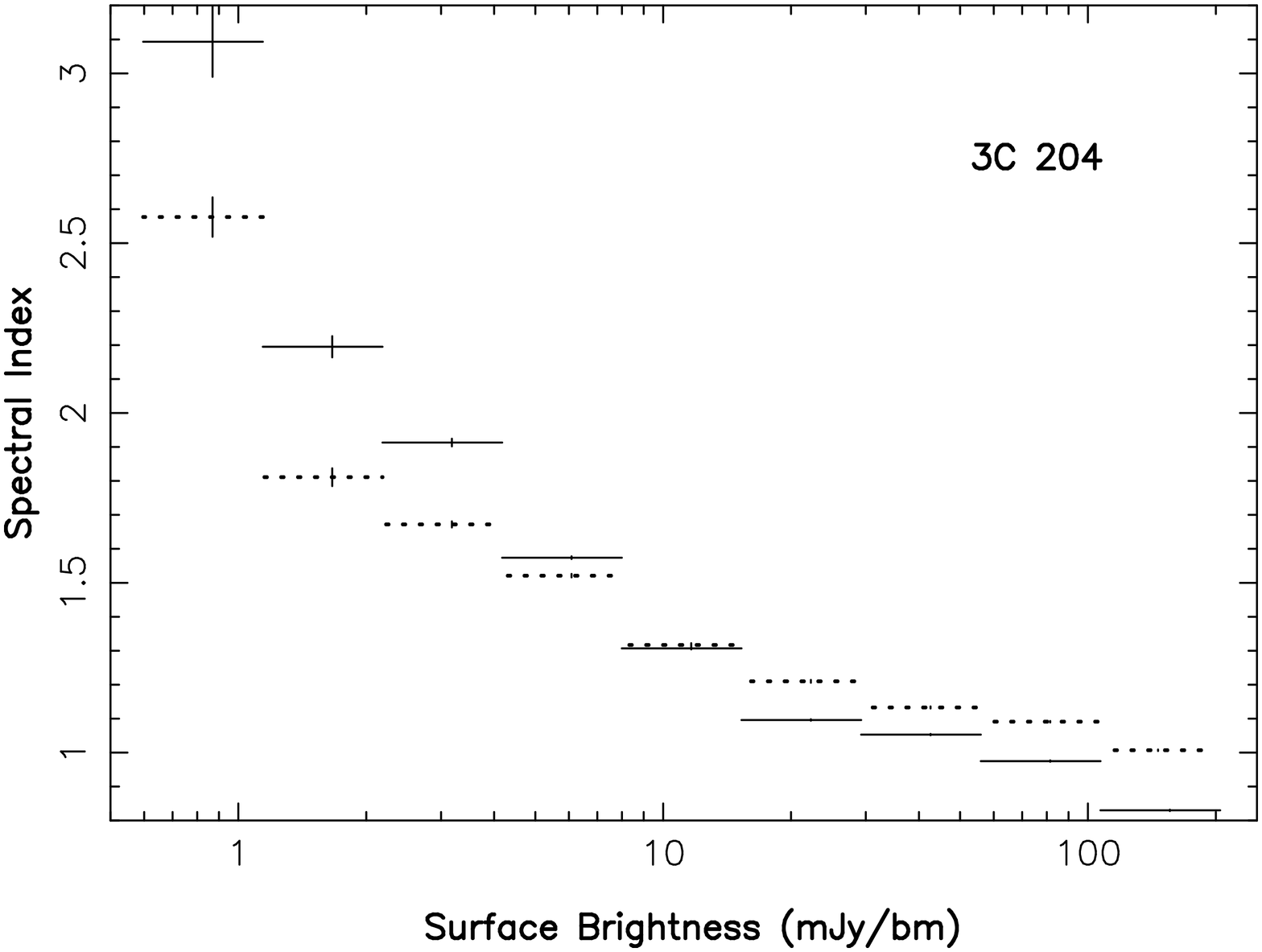,angle=-90,height=4.6cm,clip=}
\psfig{figure=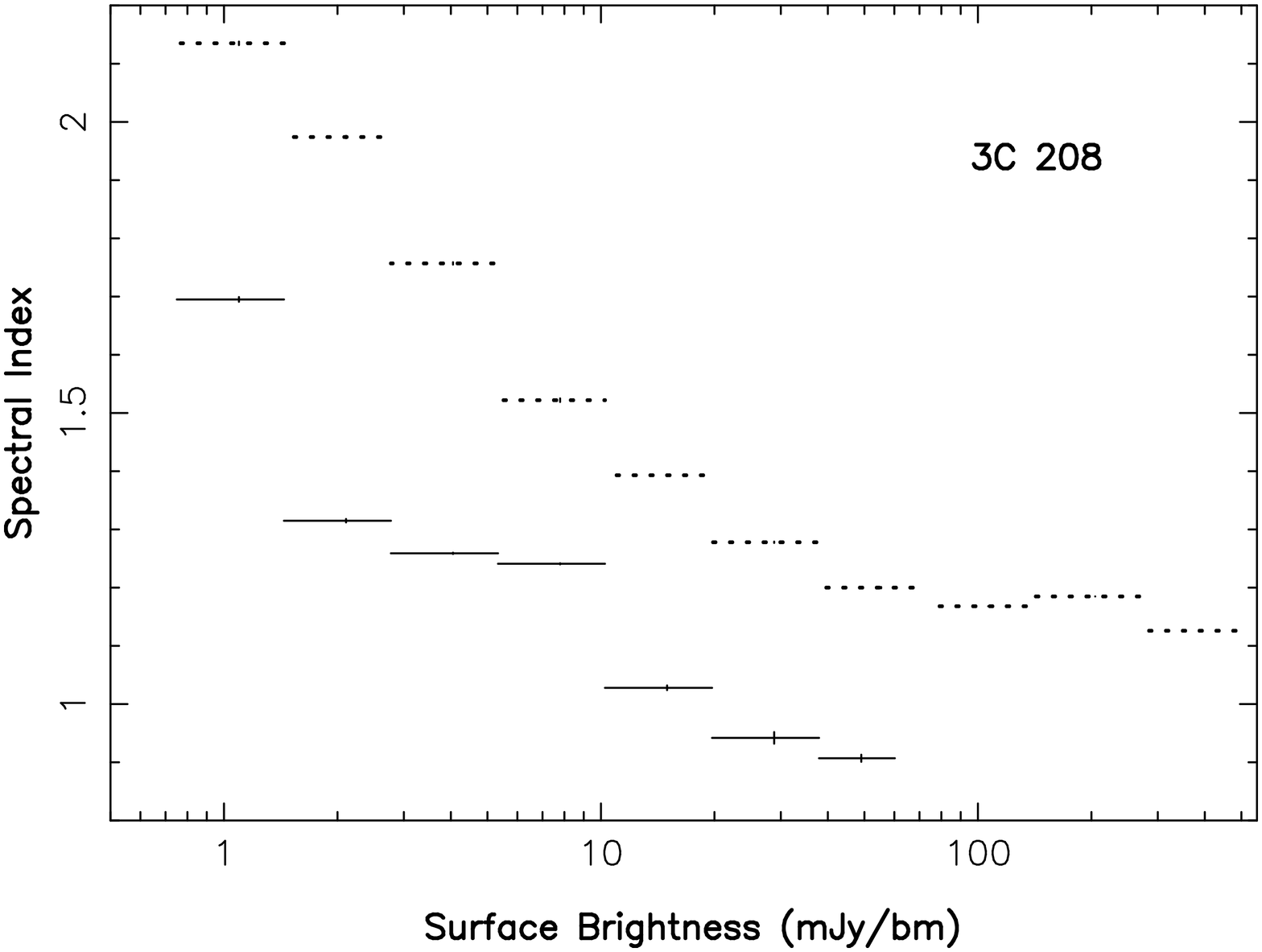,angle=-90,height=4.6cm,clip=}
}
\centerline{
\psfig{figure= 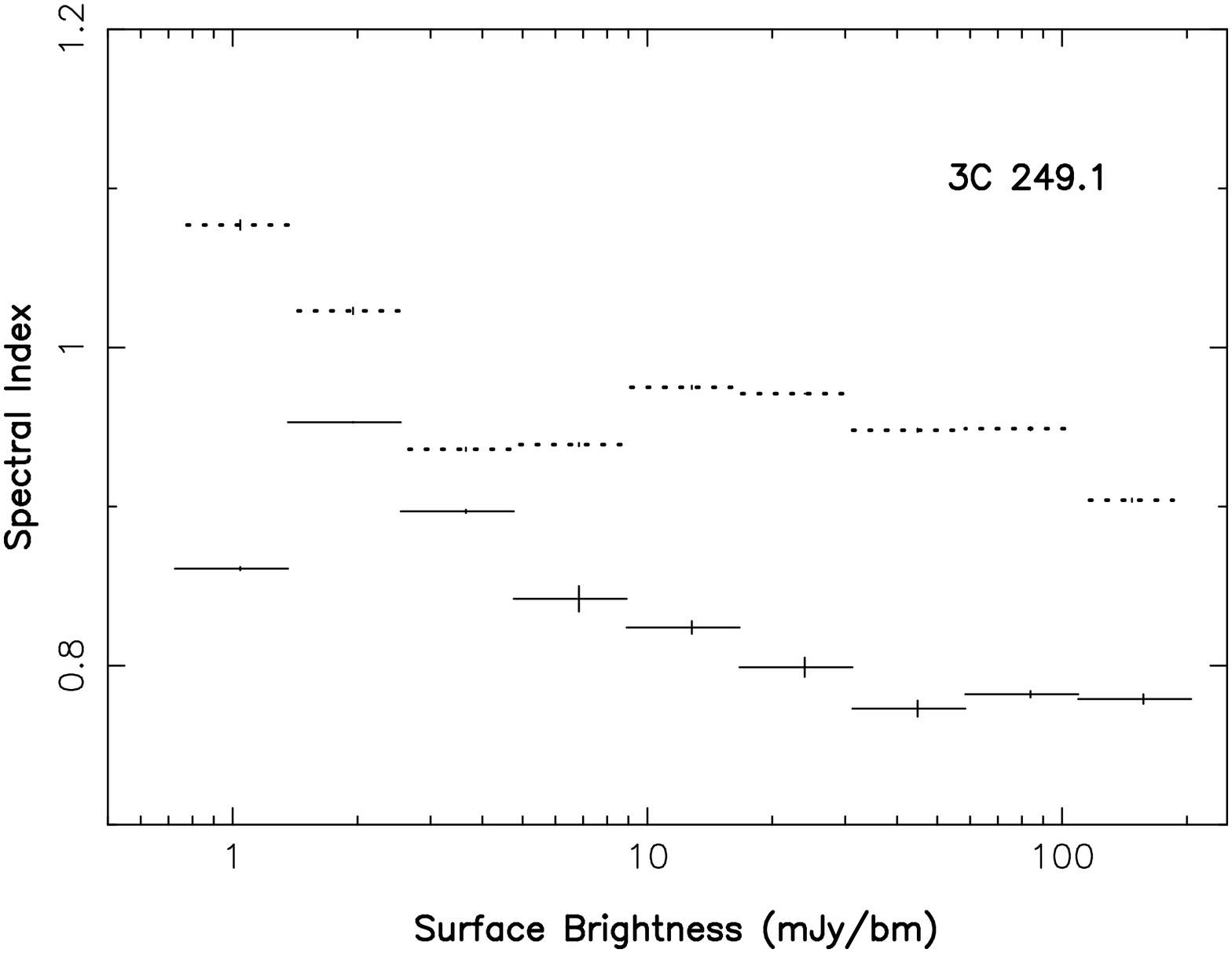
,angle=-90,height=4.6cm,clip=}
\psfig{figure=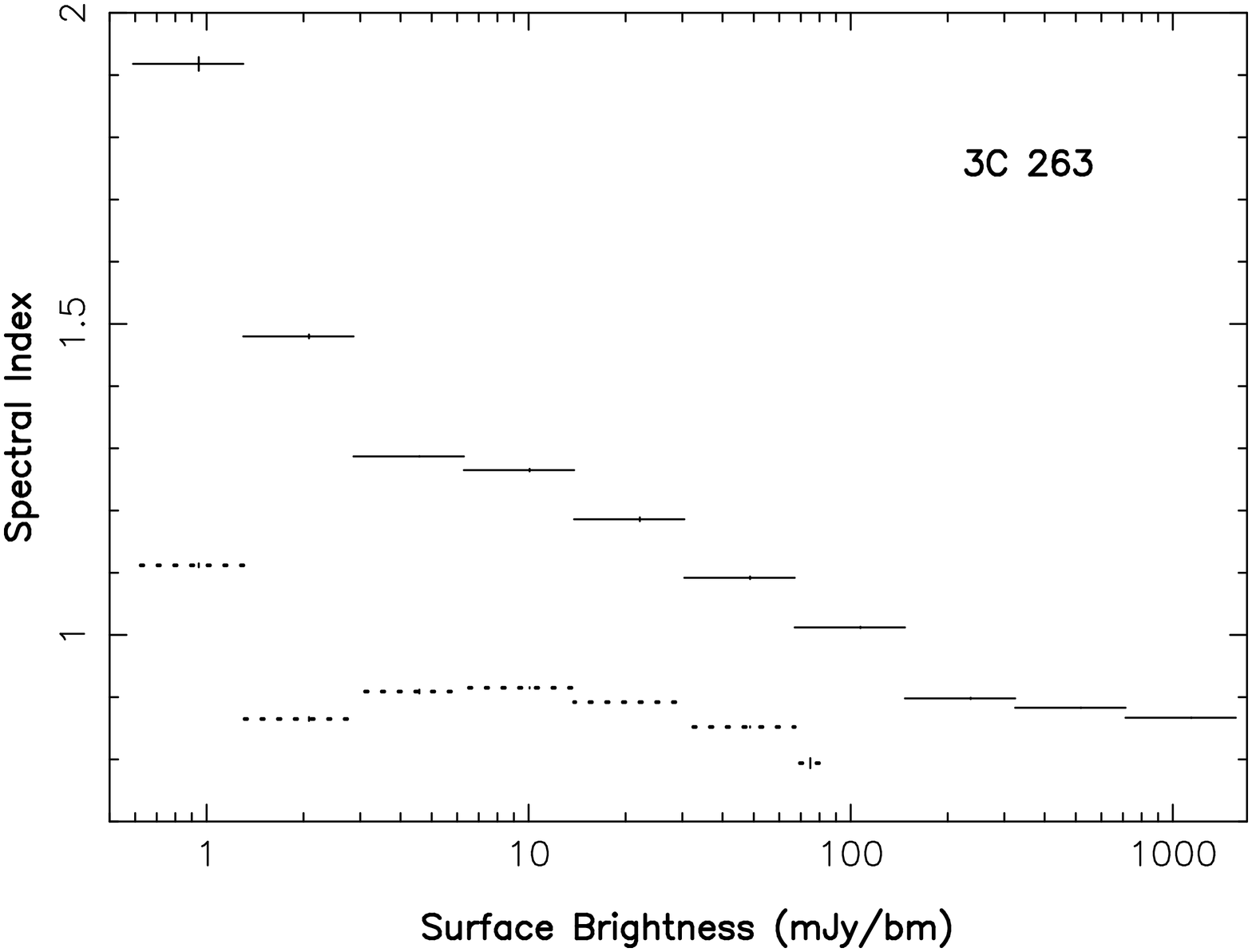 ,angle=-90,height=4.6cm,clip=}
}
\centerline{
\psfig{figure= 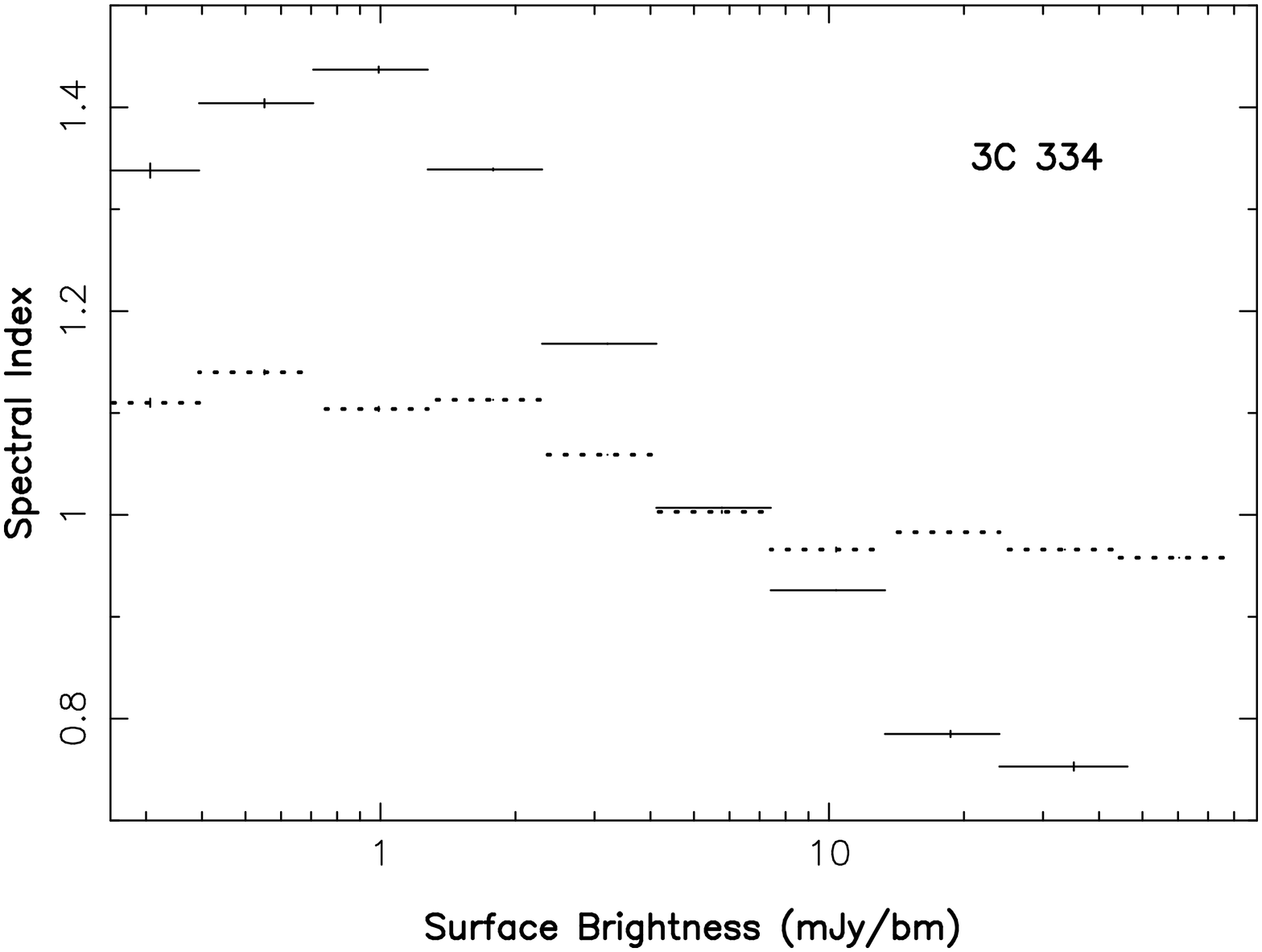
,angle=-90,height=4.6cm,clip=}
\psfig{figure=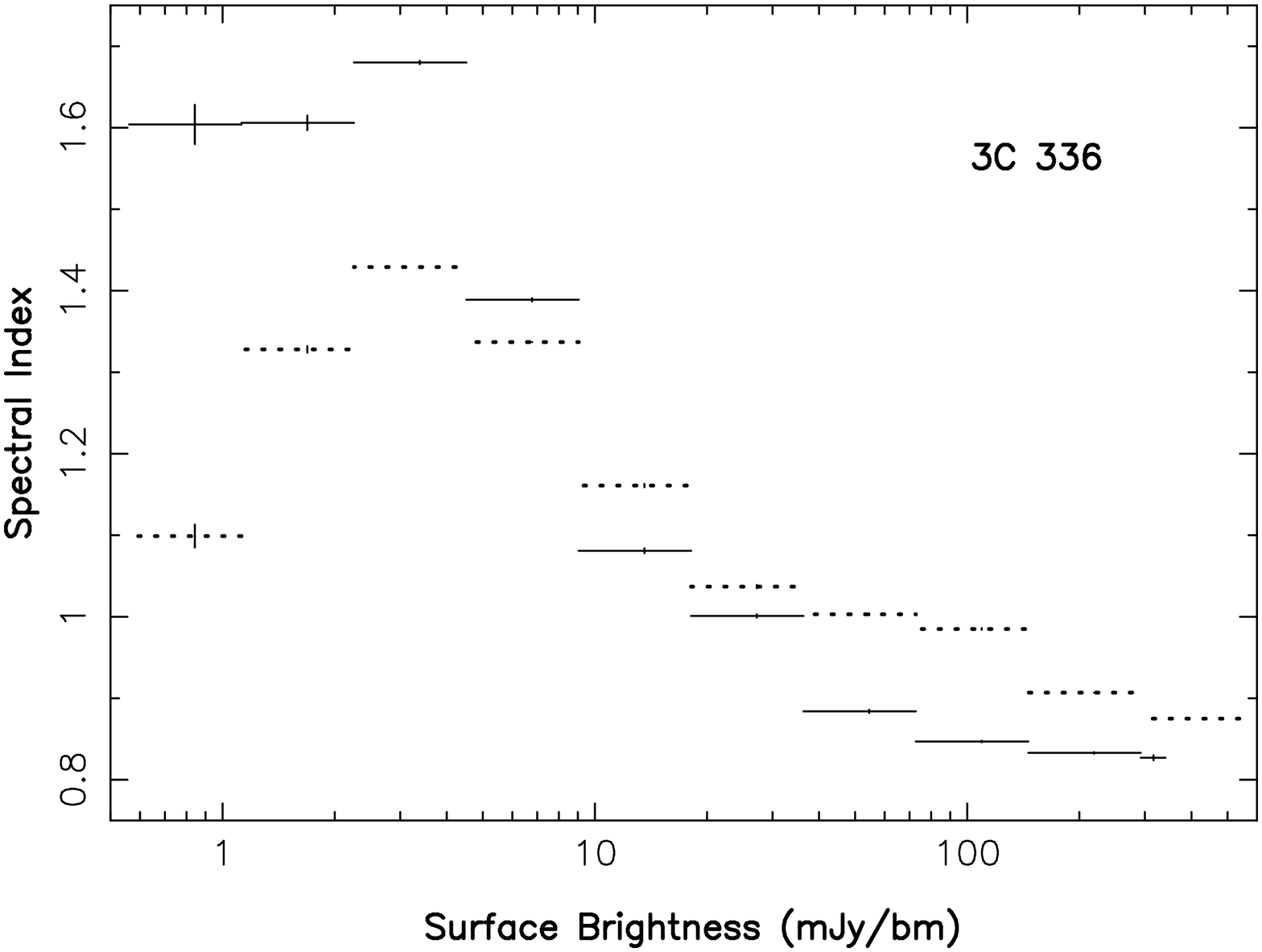 ,angle=-90,height=4.6cm,clip=}
}
\centerline{
\psfig{figure= fig3-i.ps,angle=-90,height=4.6cm,clip=}
\psfig{figure= 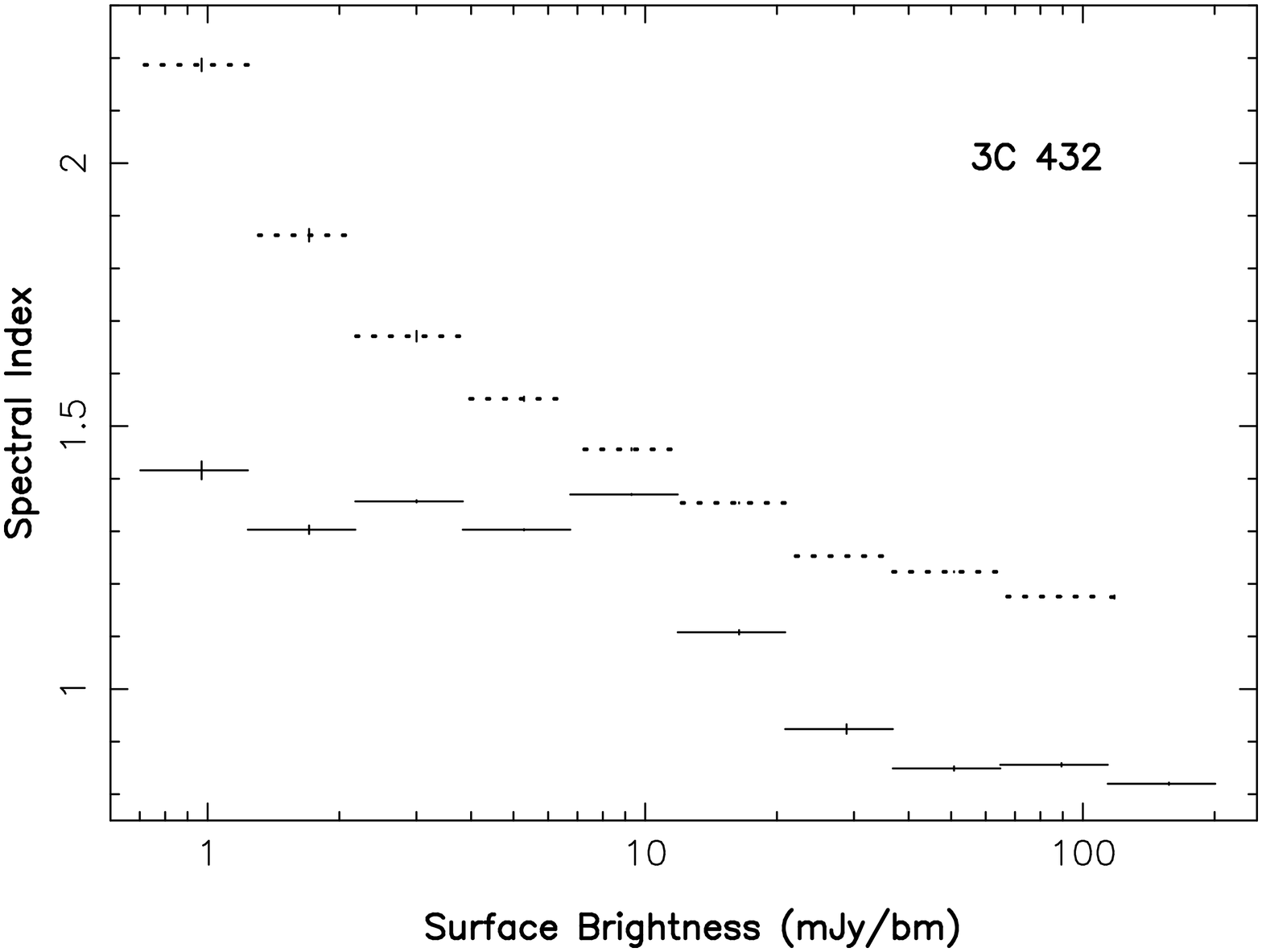,angle=-90,height=4.6cm,clip=}
}

\caption{Plots of 1.4/1.7 to 5.0~GHz spectral index versus surface brightness
for the jet and the counter-jet side of each source in the sample. Counter-jet
side: dotted lines. Jet side: continuous lines.}
\label{plots}
\end{figure*}

Fig.~\ref{plots} shows the spectral indices for both sides of each
source plotted against surface brightness. In each diagram the length
of a horizontal line shows the range of surface brightness spanned by
the zone. The vertical lines indicate the formal errors due to the
noise levels in the images at both frequencies. 

The systematic errors in spectral index are less well determined.
Small errors in the reconstruction algorithms were investigated by
making images by a `maximum entropy' algorithm ({\sc VTESS}), or by
using different parameters in the {\sc CLEAN} algorithms. These tests
typically indicated an uncertainty comparable to the errors due to the
noise in the regions of highest surface brightness, increasing to $\sim
0.1 - 0.3$ in the region of lowest surface brightness. In all cases
however, the effect was, as expected, the same for both the jet side
and counter-jet side, thus not affecting our conclusions.

\begin{figure}
%%\vspace{12cm}\centerline{
\centerline{
a.\psfig{figure=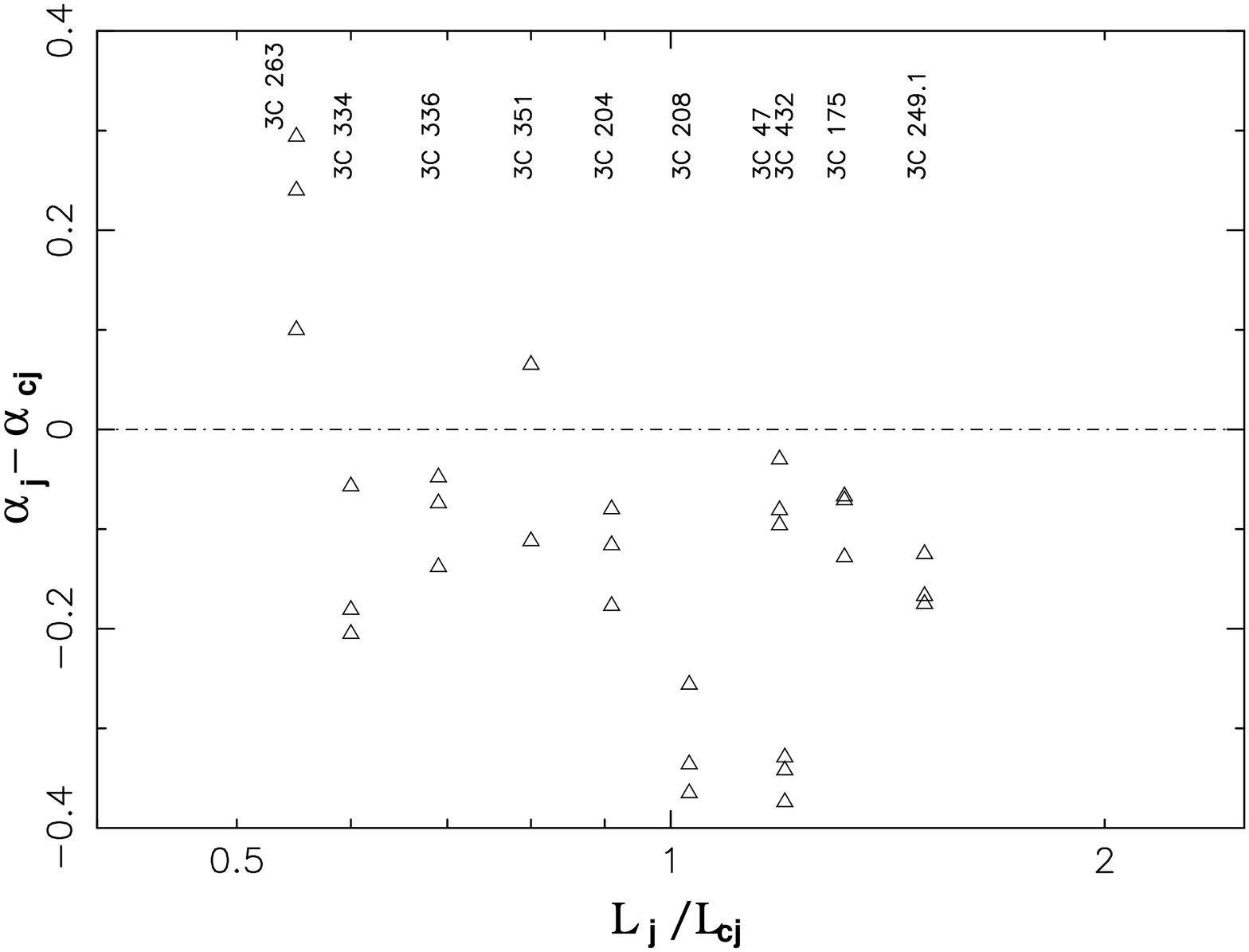,angle=90,height=6cm,clip=}}
\centerline{
b.\psfig{figure=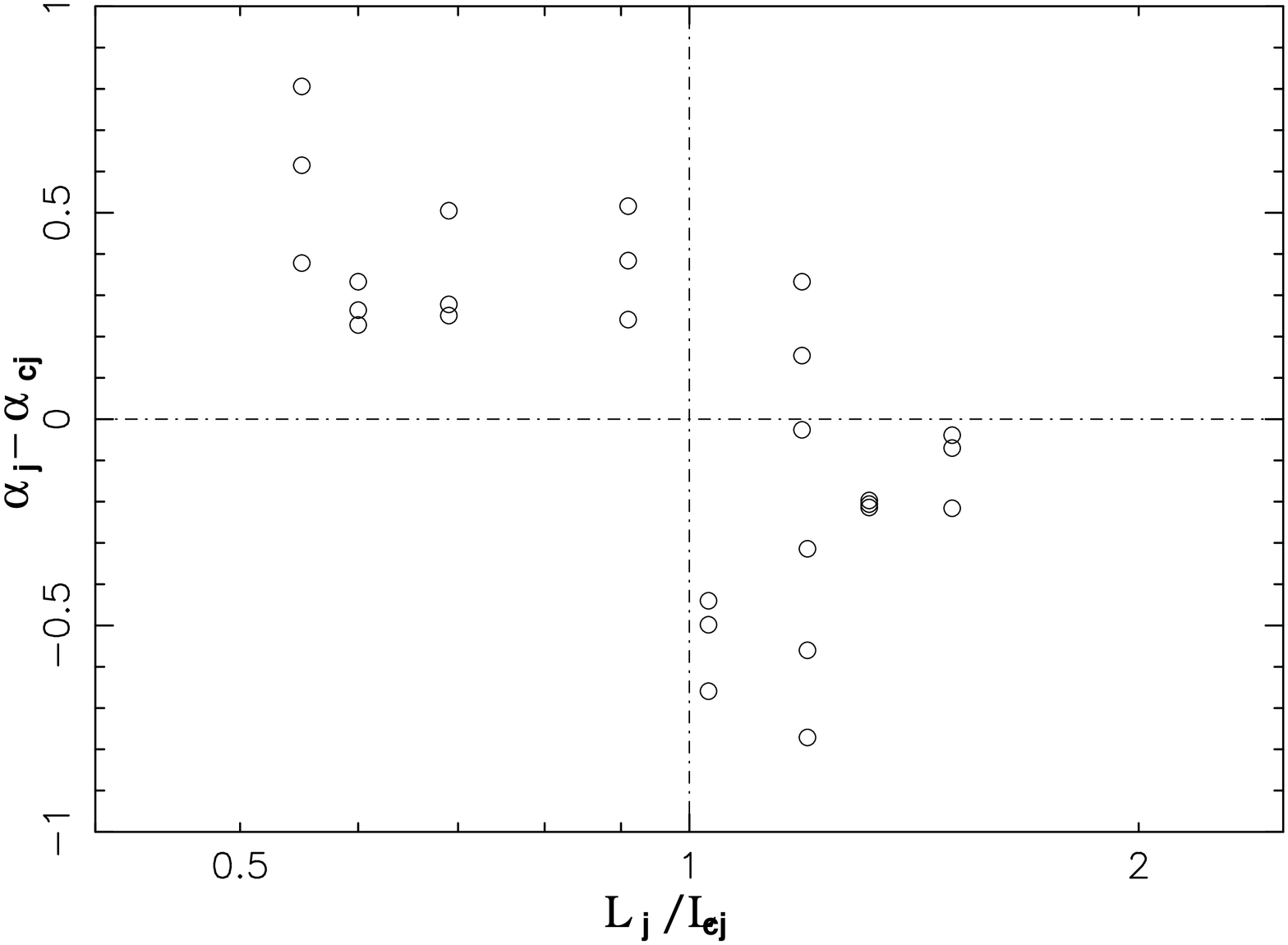,angle=90,height=6cm,clip=}}
\caption{The spectral indices of low and high surface brightness
regions related to jet side and lobe length. Figs.~\ref{hilo}a and
\ref{hilo}b show spectral index differences between the jet and
counter-jet lobes of each source for the highest and lowest three
surface brightness bins, respectively. The spectral indices of the jet
and counter-jet lobes are $\alpha_j$ and $\alpha_{cj}$, respectively, and
their lengths are L$_j$ and L$_{cj}$.}
\label{hilo}
\end{figure}

With the single exception of 3C263, the plots show that the high
surface brightness regions have flatter spectra on the jet side.  The
spectral differences in the low-brightness regions show no obvious
pattern in these plots alone, but we noticed that 3C263, the one
source which breaks the rule at high surface brightness, is strikingly
asymmetric. This led to the discovery that the spectral index
difference in the low-brightness regions correlates strongly with the
length of the lobe: the longer lobe has the flatter spectrum. The
extent to which our data support these assertions is illustrated in
Fig. \ref{hilo}. In this figure the lobe length ratio has been
calculated using the distance from the core to the furthest 3$\sigma$
contour. (Using the definition of distance from the core, through the
hotspot to the the furthest 3$\sigma$ contour 3C249.1 has a ratio of
jet-side to counterjet-side lengths of 0.6, but the figures for all
other sources are only slightly changed, and no other sources show a
change in the sense of the asymmetry.)

Fig. \ref{hilo}a shows that for all sources, except 3C263, the jet
side spectra in the highest surface brightness regions of the source
are consistently flatter than their counterparts on the counter-jet side
(i.e. they fall below the dotted line). As some of the sources have a
substantial peak brightness asymmetry the highest bins are often found
only on one side. To create properly paired highest bins we adopted the
following procedure: on both sides we summed all the
flux above the lower surface brightness bound of the top bin on the
side with the lower peak surface brightness. As can be seen from inspection of
Fig. \ref{plots}, because of the magnitude of the differences, other
sensible ways of pairing the bins would have left the conclusions
unchanged. 3C351 has a large peak surface brightness asymmetry, and
only four paired surface brightness bins. For this source we have
included only the two highest surface brightness bins in
Fig.~\ref{hilo}.

Fig. \ref{hilo}b shows a strong dependence of the spectral
asymmetry in low-brightness regions, not on jet side, but on length of lobe. 
This is evident from the fact that nearly all the points lie in the second
and fourth quadrants of the diagram. Thus when the jet side is longer 
(right of dotted line), the jet side has a flatter spectrum 
(below dotted line), but when it is shorter it has a steeper spectrum. 
The flatter spectrum is found on the longer side in all sources except 3C47.
 
The probability that at least 9 out of 10 sources have flatter spectra 
on the jet side by pure chance is 1\%. (For a two-tailed distribution, i.e. 
either $\geq$ 9/10 or $\leq$ 1/10, the probability is 2\%.) The formal statistical 
significance is not great, but we note that in the one exception, 3C263, 
the jetted lobe is very much shorter than the unjetted lobe, so the 
factors related to lobe length may well dominate in this source. If we exclude 
3C263, the probablity for 9/9 sources is 0.2\% (0.4\% 2-tailed). The 
probability that 8 out of 9 sources follow the lobe-length relation of 
Figure 4(b) by pure chance is 2\% (4\% 2-tailed). These probabilities follow 
directly from the binomial distribution. 

For completeness, we also note that the \lang effect for our sample is
found in 7/10 of our sources: the exceptions are two sources showing
very little asymmetry (3C263 and 3C249.1) and one (3C204) apparently
showing the reverse effect (although the source is almost completely
depolarized by 1.5~GHz so measurement of the asymmetry is difficult).

\section{Interpretation}
In this section we discuss various attempts to explain the two correlations 
presented in the previous section, and eliminate as many as we can. 

\subsection{Jet side and $\alpha$ in bright regions}

The hotspot on the approaching (jet) side may make a larger contribution
to the spectral index of the high-brightness regions than its
counterpart in the receding lobe as a result of Doppler beaming (e.g.
Tribble 1992); this is perhaps the simplest model for a sytematically
flatter spectrum on the jet side. However, the simplest models of this
type fail to explain the present observations, in which regions of the
same surface brightness were compared. (This applies to all models based
on different ratios of flat- and steep-spectrum power-law components,
regardless of how difference in ratio is achieved.) Consider a model
with a steep spectrum background of uniform surface brightness on which
are superposed flat-spectrum hot-spots of different intensities (and
shapes) on the two sides: points with the same suface brightness will
then have the same spectrum. (This is true regardless of angular
resolution.)  Certain more complicated models may produce the desired
effect; for example, a background symmetric about the nucleus, peaking
under similar-sized hot-spots, would produce a flatter spectrum at the
brighter hot-spot when comparing regions of equal surface brightness.
However, modifying the symmetry and shape of the background
distribution, or the relative sizes of the hotspot will not generally
give the same result.

Another way to obtain a spectral index asymmetry in a
relativistically-expanding source would be to postulate that the
intrinsic spectrum is curved and that the asymmetry is produced by the
Doppler shift.  If the spectral index increases with frequency, and
the approaching lobe is seen at a significantly lower emitted
frequency, an asymmetry of the right sense could be produced.  This
mechanism could operate in smaller regions with significant flow
velocities such as hotspots.    A calculation using
theoretical `single burst' spectra with injection spectral index of
0.5 shows that a velocity of $ \approx 0.3c$ at 30$^\circ$ to the line
of sight (and correspondingly smaller speeds at smaller angles) is
enough to cause the observed effect, if the relativistic component
dominates the hotspot spectrum, and the curvature is due to synchroton
ageing of an initial power-law electron population. The less curved
`continuous injection' spectra require somewhat higher speeds to
explain the correlation. Observational evidence for curved hotspot
spectra at high resolution comes from Cygnus A \cite{car91}, but
adequate resolution over a large frequency range is available for few
other sources \cite{mei89}, and in particular for none of our sample.

\begin{figure}
%%\vspace{6cm}
\centerline{
\psfig{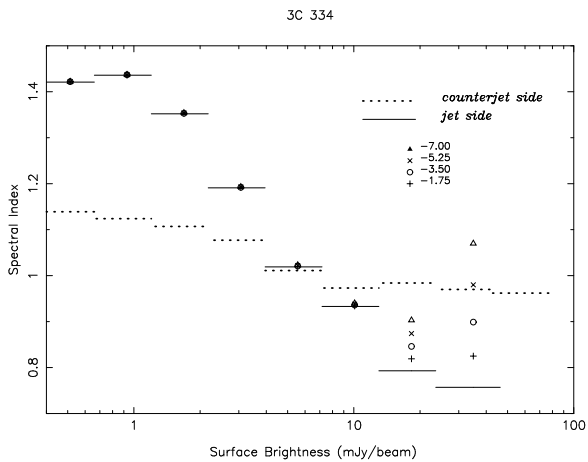}}
\caption{The effects of removing a flat spectrum component from the
jet-side hotspot in 3C334. The symbols indicate results after
subtraction of 5~GHz flux. Horizontal binning bars are omitted for
clarity; the flux subtracted is indicated in mJy in the legend. As
progressively more flux is subtracted from the jet side hotspot, its
spectrum necessarily becomes progressively steeper, but the
intermediate surface brightness regions on the jet side still have
flatter spectra.}
\label{hsremove}
\end{figure}

We have tested the hypothesis that the spectral differences between
high-brightness regions are due to high flow velocities in the hotspots in two
ways, as follows:
\begin{enumerate}
\item Table~\ref{tab:flux} lists the flux densities of the hotspots in
our sample. We present two sets of hotspot flux densities, that of
BHLBL (obtained by a well-defined, though arbitrary set of rules
defining a hotspot) and our own estimates (for which the hotspots were
chosen subjectively).  In most cases our estimate of total hotspot
flux density is close to that of BHLBL, but in a few cases there is
obvious disagreement, as on the counter-jet side of 3C432 where
different features have been chosen in the two cases. In the majority
of the sources there is a good case for asserting that the jet-side
hotspot is the brighter of the two, but there are striking exceptions
--- 3C175, 3C208, 3C336 --- whichever set of hotspot criteria is
used.  In such sources the flatter spectrum on the jet side obviously
cannot be blamed on a greater flat spectrum contribution from the
hotspot.  Thus, although relativistic flux--boosting of a
flat--spectrum hotspot in a steeper spectrum lobe could potentially
account for the effect, it clearly cannot do so in all cases. The
inclusion of two sets of flux density estimates emphasises that these
conclusions are robust and insensitive to details of hotspot definitions.

\begin{table}
\vspace{5mm}
\caption{5 GHz flux densities of hotspots in mJy.}
\begin{tabular}{lrrp{0.0cm}rrrr} \hline
 
Source & $S_j^{tot}$ & $S_{cj}^{tot}$ &&$S_j^{tot}$ & $S_{cj}^{tot}$  &$S_j^{peak}$  &$S_{cj}^{peak}$ \\
\hline
 & \multicolumn{2}{c}{BHLBL} &&  \multicolumn{4}{c}{our estimates}\\
\hline 

3C47   & 268 &  -  & &257 &  - & 198& 25\\
3C175  &  64 & 152 & &62  & 138& 25 & 65\\
3C204  &  61 &  42 & & 60 & 40 & 43 & 30\\  
3C208  &  28 & 203 & & 36 & 198& 21 &139\\
3C249.1&  86 & 188 & & 80 &120 & 48 & 18\\
3C263  & 528 &  21 & &500 & 23 &330 & 13 \\
3C334  &  20 &  -  & & 18 & -  & 6  &  3 \\
3C336  &  95 & 254 & & 95 &280 & 50 & 40 \\
3C351  & 201 &  -  & &202 &  5 &156 &  1\\
3C432  & 105 &   6 & &103 & 280& 73 & 34\\
\end{tabular}
\label{tab:flux}
\end{table}

\item The two effects of high flow velocity (the Doppler effect
shifting a curved spectrum in frequency and the relativistic flux-boosting
of the approaching flow) may be thought of as increasing the
proportion of flat-spectrum component in the jet-side hotspot.  Even
if the fast flow is confined to the hotspot, it might be
suspected that, at the resolution used for the spectral-index
analysis, beam smearing could spread a flatter spectrum from the
hotspot to adjacent regions of somewhat lower surface brightness.

To test the possibility that the effect is solely due to a single
compact hotspot, we used the three sources in which the spectral index
difference $\alpha_j -\alpha_{cj}$ reverses between high- and
low-surface-brightness regions (3C204, 3C334 and 3C336). If the above
suspicion were well-founded, artificial spectral steepening of the
jet-side hotspot spectrum alone should remove the correlation between
jet side and flat spectrum. The jet-side hotspot was modelled using
the 2D Gaussian fit to the high resolution image of the hotspot at
5~GHz made by BHLBL. This fitted hotspot was convolved down to
the resolution of our spectral analysis.  This model was then used to
remove 5~GHz flux progressively from the site of the high resolution
fit to the hotspot. Even when the amount of flux subtracted from the
hotspot was large enough to produce a spectrum of the compact
component which was steeper than that on the counter-jet side, there
still remained a region of intermediate brightness which was largely
unaffected by the changes to the hotspot, and which retained its
flatter spectrum.  Fig. \ref{hsremove} shows the results for a typical
source (3C334). Thus it is clear that ``contamination'' of the
intermediate surface brightness regions by an insufficiently resolved
flat spectrum hotspot is not enough on its own to account for the
observations.
\end{enumerate}

\begin{figure}
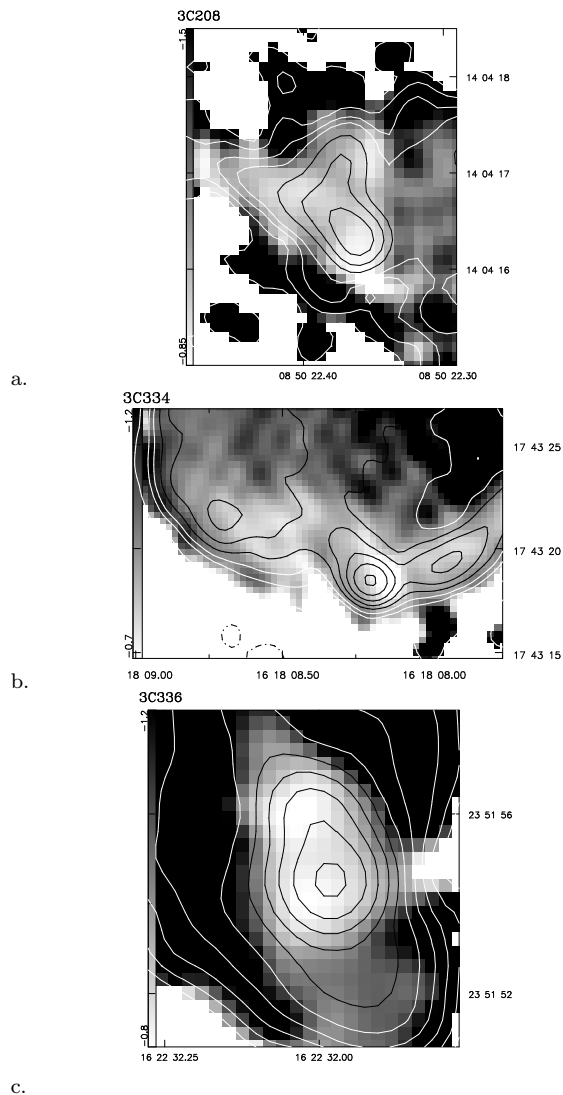

%%\vspace{14cm}
a.\centerline{
\psfig{figure=fig6-a.ps,angle=0,height=5cm,clip=}}
b.\centerline{
\psfig{figure=fig6-b.ps,angle=0,height=4cm,clip=}}
 \centerline{
\psfig{figure=fig6-c.ps,angle=0,height=5cm,clip=}}
c.
\caption{Detail of jet side lobes of 3C~208, 3C~334 and 3C~336. Grey
scale representations of spectral index are superposed on total intensity contours.}
\label{new}
\end{figure}

We conclude that, if relativistic motion is responsible for the
spectral index asymmetry, then the fast forward flow is not confined
to a single compact hotspot. What regions, then, might take part in
this forward motion? It is well known that many sources have double or
even multiple hotspots (3C351 is a striking example, and in the
present sample 3C175, 3C334, 3C336 and 3C432 show double hotspots at
higher resolution), but that by itself does not explain why there
should be fast forward motion in both.  One possibility is that the
point of impact of the jet on the shocked intergalactic medium is
recessed (i.e. not at the extreme end of the lobe) and fast flow
continues beyond the initial hotspot, as the images of 3C204, 3C334
and 3C336 might indicate (Fig. \ref{maps}).  Inspection of the
spectral index image for 3C334 shows that a ridge with relatively flat
spectrum indeed extends along the ridge of \hsb beyond the hotspot
(Fig.  \ref{new}b). At higher resolution (BHLBL), the hotspot of 3C336
(single in our images) divides into two compact bright features. These
coincide with two flatter-spectrum regions in Fig. \ref{new}c. 3C208
also has a recessed hotspot on the jet side, but in that case the
flattest spectra occur on the hotspot and on a region extending
northward from it (Fig. \ref{new}a), at about the same distance from
the quasar. This suggests that the jet has split before reaching the
hotspot, or perhaps that one of the flat-spectrum regions represents
an earlier hotspot, now detached from the jet but still being fed with
high-speed material, as in the numerical simulations of Cox, Gull \&
Scheuer \shortcite{cox91}. It is also possible that projection effects
have transformed a gentle bend in the flow into a sharp ($>90^\circ$)
change of direction, but that would imply a very small angle of source
axis to line of sight, and a correspondingly large true aspect ratio
for the source. Like 3C334 and 3C336, 3C47 and 3C175 show relatively
flat spectra along \hsb ridges extending out of the jet-side hotspots,
but in these sources the flow seems to turn through a (projected)
right angle at the hotspot. Perhaps these flows have more in common
with what is going on in 3C208 than with flows in 3C334  and 3C336.

Numerical simulations of radio emission from hotspots generated by
relativistic jets have so far been restricted to the axisymmetric
case.  Komissarov \& Falle \shortcite{KF96} find that the brightness
distribution of the approaching hotspot is dominated by the conical
termination shock, whilst the emission from the receding hotspot comes
mainly from the high-pressure part of the backflow, and is therefore
limb-brightened. High-resolution non-relativistic simulations of
Norman \shortcite{Norm96}, suggest that there is a region of
supersonic turbulence near the leading edge of the source (extending
4--5 lobe radii in these simulations) in which the jet is violently
deflected. The jet can therefore impinge obliquely on the contact
discontinuity to produce a deflected or wall jet
\cite{WG85,W89,cox91,NB92}.  These deflected flows are morphologically
very similar to the flat-spectrum features observed in sources such as
3C334. A key feature of these simulations, from our point of view, is
that they lead to oblique shocks and therefore to relatively large
forward velocities, so that significant beaming can occur over much of
the post-shock flow, especially when the hotspot is recessed. 

The simulations
also indicate that the observations of the approaching and receding
hotspots may well be dominated by different parts of the flow, and
might then have different spectra. Observationally, the more diffuse
appearance of the counterjet-side hotspots also supports this idea.

\subsection{The length of the lobe and $\alpha$ in lower brightness regions}
 
 The expansion velocities of the low-brightness regions are unlikely
to exceed a few per cent of $c$ \cite{Sch95}, so it is difficult to
see how beaming effects could account for large-scale spectral
asymmetries, even if the intrinsic spectra are curved.  Two obvious
possible causes of the correlation of the continuum radio spectra with
the size of the radio lobe are synchrotron loss and adiabatic
expansion. We consider these in turn below.

\begin{enumerate}

\item Synchrotron losses 

Other things being equal, the break frequency due to synchrotron loss
varies as $B^{-3}$ ($B$ = magnetic flux density). The equipartition
estimate of $B$ varies as (linear size)$^{-6/7}$, for given radio
power. If, on the other hand, the magnetic field in the larger lobe is
simply a homologously expanded copy of that in the smaller lobe, then
$B \propto({\rm linear \; size})^{-2}$.  In either case, the break
frequency depends sensitively on linear size, and this dependence
would produce a higher break frequency (and hence a flatter spectrum
over a fixed frequency interval) in the larger lobe, i.e. a
correlation in the sense that is observed. The theoretical predictions
are more complicated if we compare regions of equal surface brightness
in the two lobes, but in essence the result remains the same.
  
Blundell and Alexander (1994) pursued this line of argument to explain
the correlation between jet side and flatter spectrum. They argued
that the near (jet) side is observed at a later stage of development
(owing to light travel time effects), and is therefore the
larger. While we cannot accept that part of their hypothesis, because
the longer lobe is on the {\em counter-jet} side in 5 of our 10
sources, the strong dependence of synchrotron loss on linear size
remains a plausible explanation for spectral asymmetries in
low-brightness regions.

\item Adiabatic losses.

The observation of spectral index gradients in the lobes indicates that 
individual regions have spectra steepening with frequency, 
possibly because of synchrotron losses (this need not conflict with a 
fairly straight spectrum for the whole source as  
many sources are dominated by radiation from hotspots where very little 
synchrotron loss has occurred at frequencies less than $\sim$ 10~GHz). We now 
ask how different amounts of expansion in the two lobes, acting on these 
curved spectra, might affect the differences between their observed spectral
indices.

To isolate the effects of expansion, consider a simple model: two
lobes were identical initially; then one expanded adiabatically by a
linear factor $R$ which is a little greater than 1. The magnetic field
in the larger lobe becomes $R^{-2}$ of the field in the corresponding
bit of the smaller lobe, and the electron energy distribution is
shifted downwards (electron energy $\propto R^{-1}$), with the result
that the entire spectrum is shifted downwards in frequency by a factor
$R^{-4}$. Thus we should expect the larger lobe to have a spectrum
that is steeper over the same frequency range. This is the reverse of
the correlation observed in Fig.~4. Closer consideration (below) modifies that
conclusion, for we must remember that we compare regions
of equal surface brightness on the two sides.

Suppose that a certain region of the smaller lobe has  
spectral index $\alpha$ and surface brightness $S/\Omega$. The corresponding 
region in the larger lobe has spectral index 
\[ \alpha'=\alpha+4\log R \frac{d\alpha}{d(\log \nu)} \]

\noindent and surface brightness given by 

\[ \log (S/\Omega)' = \log (S/\Omega) - 4(1+\alpha) \log R.\]

Therefore a region of surface brightness $S/\Omega$, in the larger lobe, is 
expected to have spectral index

\begin{eqnarray}
\alpha'' &=&\alpha'+4(1+\alpha)\log R \frac{d\alpha}{d(\log S/\Omega)} \nonumber\\
  & = & \alpha + 4 \log R \left(\frac{d\alpha}{d(\log \nu)} 
   + (1+\alpha)\frac{d\alpha}{d(\log S/\Omega)} \right) 
\end{eqnarray}

The last term is negative: brighter patches of source have flatter spectra. 
Thus we expect the larger lobe to have a flatter spectrum {\em for regions 
of equal surface brightness} if 
\[  \left|(1+\alpha)\frac{d\alpha}{d(\log S/\Omega)} \right| 
    > \frac{d\alpha}{d(\log \nu)} \]

If we assume that the spectral gradient along the source is due to
synchrotron losses we can evaluate the importance of adiabatic loss as
an explanation for the observed correlation. We can calculate
$\frac{d\alpha}{d(\log \nu)}$ as a function of $\alpha$ from the
synchrotron spectrum of a given theoretical electron energy
distribution. The results are shown in Fig. \ref {func}, which shows
the results for two values of injection index for the case of a sharp
energy cut-off. Another energy distribution of interest, that with no
pitch--angle scattering of the electrons \cite{Kardashev}, follows the
above case closely, until $\alpha$ approaches
$\frac{4}{3}\alpha_{injection} + 1$, which is the maximum spectral
index possible in this model, so that $\frac{d\alpha}{d(\log
\nu)}\rightarrow 0.$ $\frac{d\alpha}{d(\log S/\Omega)} $ has to be
estimated directly from our observations, as it involves the whole
histories of synchrotron losses in different parts of the source.

The results are illustrated in Fig. \ref{func}. One point is plotted for
each source. The value of $\alpha$ is taken as the mean of the three
lowest surface brightness bins in the larger lobe.  Errors in
%%${d\alpha}/{d(\log S/\Omega)}$ were estimated from fitting the slopes,
%%and those in 
$\alpha$ are taken as the difference between the maximum
and mean values of $\alpha$ in the lowest three bins.

The points need to fall above the line for adiabatic expansion to
explain the observed correlation. It can be seen that this is indeed
the case for most of the sources and two reasonable models of
$\frac{d\alpha}{d(\log \nu)}$.  We conclude that adiabatic expansion
may contribute to the observed correlation.

\begin{figure}
%%\vspace{6cm}
\centerline{
\psfig{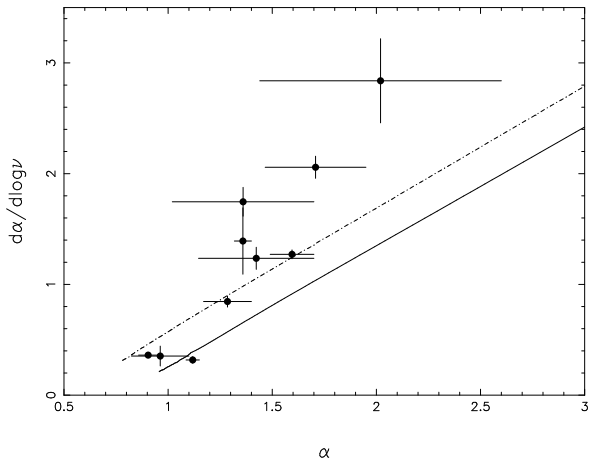}}
\caption{Calculated values of $\frac{d\alpha}{d(log\nu)}$ and observed
values of
\mbox{$(\alpha+1)|\frac{\Delta\alpha}{\Delta\log(S/\Omega)}|$}
plotted on the same scale. The lines are $\frac{d\alpha}{d(log\nu)}$
calculated for energy spectra with a sharp energy cut--off, and
injection spectral indices of 0.5(dotted line) and 0.75(solid
line). One observed point, from the longer lobe, is plotted for each
source.}  
\label{func}
\end{figure}

It seems desirable to check the prediction that the larger lobe {\em as a 
whole} has the steeper spectrum, but it is not clear how to do so. 
The difficulty is to identify which material is potentially affected by
orientation-dependent effects.  We have shown that such effects are not
restricted to easily-identifiable regions such as hot-spots, and therefore 
that setting an intensity threshold will not unambiguously separate them.
 
\end{enumerate}

\section{Conclusions}

In a sample of ten high-powered radio quasars we have found the
following correlations:
\begin{enumerate}
\item In regions of high surface brightness the radio spectrum is flatter on
the jet side (9/10)
\item In regions of low surface brightness the radio spectrum is
flatter on the long side (8/9)
\end{enumerate}

If jet sidedness is a manifestation of relativistic flow, as is
commonly believed, then the strong correlation (9/10 sources) found
here between the spectral index of the hotspot and the jet side
indicates that the spectral difference between the two sides is also a
relativistic effect. 
  We have also demonstrated that the correlation of flatter
spectrum with jet side is not confined to the hotspot.  Nevertheless,
the correlation is presumably an orientation-dependent phenomenon, and
we conclude that forward motion at a significant
fraction of $c$ also occurs in regions less conspicuous than that
normally selected as `the hotspot'.

We have argued that the observed correlation cannot result from a
flat-spectrum component whose contribution to the hotspot is enhanced
by forward beaming. Doppler frequency shifting of a curved hotspot
spectrum might be able to explain the correlation. This would require
flow speeds in excess of $0.3c$. A further possibility is that
we are viewing different physical parts of the flow on the two sides as a
result of Doppler enhancement and suppression of the post-shock flow.

The correlation of lobe length and radio spectrum has yet to be fully
understood. One simple explanation is that it is due to differences in
synchrotron losses in lobes of different volumes. Expansion losses may
also play a role. The spectral asymmetry in the low surface brightness
regions can be interpreted as an environmental effect due to the
surrounding ambient gas containing the lobes. In this context, we note
that the one source (3C~263) which disobeys the high surface
brightness correlation also shows surprisingly little depolarization
asymmetry. This and the fact that the jet side is only $\sim$ half the
length of the counter-jet side indicates that the source may be
confined by a denser medium on the jet side. In this case it may be
that environmental effects produce the spectral-index and
depolarization behaviour. Hall et al. (1995) indeed find evidence for
an X--ray emitting clump associated with the short lobe of this
source. 
 
A similar study of a sample of nearby radio galaxies (in
which environmental effects are expected to dominate) is now under way.

\subsection*{ACKNOWLEDGEMENTS}

JDT, RAL and PAGS would like to thank the NRAO for hospitality. JDT thanks
the British taxpayers for their assistance in the form of a PPARC
studentship. The NRAO is a facility of the National
Science Foundation, operated under cooperative agreement by Associated
Universities, Inc. MERLIN is a national facility operated by the
University of Manchester on behalf of PPARC.

\end{document}